\documentclass[manuscript,screen,nonacm]{acmart}
\renewcommand{\keywords}[1]{}

\usepackage[ruled, lined, linesnumbered, commentsnumbered, longend]{algorithm2e}

\usepackage{tikz, tikz-layers} 
\usetikzlibrary{positioning, fit, calc}
\usetikzlibrary{shapes.geometric}
\usetikzlibrary{backgrounds}
\usepackage{calc}
\usepackage{relsize}

\newcommand{\figref}[1]{\figurename~\ref{#1}}
\newcommand{\algoref}[1]{Algorithm~\ref{#1}}

\newcommand{\secref}[1]{Sect.~\ref{#1}}

\AtBeginDocument{%
  }

\setcopyright{acmcopyright}
\copyrightyear{2022}
\acmYear{2022}
\acmDOI{XXXXXXX.XXXXXXX}

\newcommand{\WRENCH}[0]{{\smaller{WRENCH}}}

\begin{document}

\title[BottleMod: Modeling Data Flows and Tasks for Fast Bottleneck \ldots]{BottleMod: Modeling Data Flows and Tasks for Fast Bottleneck Analysis
}


\author[A. Lößer]{Ansgar Lößer}
\email{ansgar.loesser@kom.tu-darmstadt.de}
\orcid{0000-0002-7627-9664}
\affiliation{
  \institution{TU Darmstadt}
  \country{Germany}
}

\author[J. Witzke]{Joel Witzke}
\email{witzke@zib.de}
\orcid{0000-0002-0831-8078}
\affiliation{
  \institution{Zuse Institute Berlin}
  \country{Germany}
}

\author[F. Schintke]{Florian Schintke}
\email{schintke@zib.de}
\orcid{0000-0003-4548-788X}
\affiliation{
  \institution{Zuse Institute Berlin}
  \country{Germany}
}

\author[B. Scheuermann]{Björn Scheuermann}
\email{scheuermann@kom.tu-darmstadt.de}
\affiliation{
  \institution{TU Darmstadt}
  \country{Germany}
}

\begin{abstract}
  In the recent years, scientific workflows gained more and more popularity.
  In scientific workflows, tasks are typically treated as black boxes.
  Dealing with their complex interrelations to identify optimization potentials and bottlenecks is therefore inherently hard.
  The progress of a scientific workflow depends on
  several factors, including the available input data, the
  available computational power, and the I/O and network bandwidth.
  Here, we tackle the problem of predicting the workflow progress with very low overhead.
  To this end, we look at suitable formalizations for the key parameters and their interactions which are sufficiently flexible to describe the input data consumption, the computational
  effort and the output production of the workflow's tasks.
  At the same time they allow for computationally simple and fast performance predictions, including a bottleneck analysis over the workflow runtime.
  A piecewise-defined bottleneck function is derived from the discrete intersections of the task models' limiting functions.
  This allows to estimate potential performance gains from overcoming the bottlenecks
  and can be used as a basis for optimized resource allocation and workflow execution.
\end{abstract}


%
%

\maketitle

\begin{acks}
  This work received funding from
  the \grantsponsor{DFG}{German Research Foundation (DFG)}{https://www.dfg.de/}, \grantnum[https://fonda.hu-berlin.de/]{DFG}{CRC 1404}: \emph{FONDA: Foundations of Workflows for Large-Scale Scientific Data Analysis}~\cite{fonda}.
\end{acks}

\section{Introduction}

Many scientific fields have a growing need for computationally intensive data analysis, from genome analysis to satellite image processing.
Such complex analysis processes are often referred to as scientific workflows.
A scientific workflow consists of multiple processing steps, so-called tasks.
Tasks can depend on the results of other tasks, creating a dependency graph between them.

Often, a single task is simply the execution of a program on one or more specific set(s) of input data, creating one or more outputs, which may then form input(s) of subsequent tasks.
Such tasks can have very different temporal I/O behavior.
An intuitive example of this is video processing:
Reversing a video file needs complete input before it can start outputting data.
Other operations work on the data sequentially in order, like re-encoding a video:
they can begin producing output while the input is still incomplete.
If a video reversal task follows after some other task $X$, then $X$ must complete before the reversal task can start with its work.
A re-encoding task in the same position within a workflow, on the other hand, could well run in parallel to $X$---at least if task $X$ outputs data before its own completion and to the extent to which (a) computation power and (b), if applicable, network capacity for the data transfer between the two tasks are available.
All these factors determine whether, for instance, pipelined processing in a specific part of a workflow is a viable option for accelerating the execution or not.

Similar patterns exist on the output side:
A video reversal task can only start when the input data is complete, but it can output continuously while the task is progressing; a task which, for instance, counts occurrences of specific patterns in a video and outputs the total number for each pattern will only be able to start outputting results after the entire video has been processed.
Other, almost arbitrary dependencies of input and output behavior over the course of execution of a task are easily conceivable, and a large variety of behaviors exists in virtually any field in which scientific workflows are applied.

Even if the behavior of individual tasks is usually reasonably simple, the interdependencies can quickly become complex.
Depending on the data input and output behaviour of predecessor and successor tasks in the workflow, very different execution behaviour and very different resource utilization over time will result.
How to understand and analyze these interdependencies given the structure of a specific workflow?
How to model them appropriately?
How to draw conclusions from them?
These are the questions we are tackling here.

We propose a way to model a task's I/O behavior over computation time in an abstract way.
We show how, based on such models, one can derive and predict the overall progress of the workflow, as well as the structure of the bottlenecks limiting the execution performance.
The models are modular and allow the combination of task models to express chains of tasks or even complete scientific workflows.
They are simple, yet flexible enough to describe a wide range of task behaviors.
And they are constructed in a way that allows for quick and lightweight computations.
The algorithm which we propose to do so operates in a quasi-symbolic way on models that are given by piecewise-defined functions.
It is inspired by discrete-event simulation and considers only those points in time where the involved functions change to the next piece.
This is a novel take on modeling the execution behavior of scientific workflows.
It scales much more favourably than any existing approach, and therefore also allows, e.\,g., for
repeated evaluation during a workflow execution, adapting the predictions to live measurements.

The resource management for workflow execution---i.\,e., an operating system in conjunction with a workflow execution environment like e.\,g., Nextflow~\cite{nextflow} with Kubernetes~\cite{kubernetes}---can make more informed decisions if it knows the bottlenecks and the potential performance gain when they are resolved.
The specific construction of a scheduler that makes use of the modeling approach proposed here is out of scope for this work.
Nevertheless, one possible application of runtime predictions from task I/O models is to determine the potential performance gain and upcoming resource demand when the resource allocation is changed.
This allows, for instance, for a comparison of different scheduling options.

We introduce our formal model for tasks and their execution in \secref{sec:math-model};
It describes the relations between input, output, and resources.
Based on this model, we analyze individual tasks and the combination of tasks to form workflows, their progress, their bottlenecks, and the actual resource utilization in \secref{sec:process-sim}.
We discuss the practical application of the approach and the challenges and tradeoffs involved in \secref{sec:impl}.
An evaluation with an example workflow confirms the model's ability to predict real-world executions in \secref{sec:eval}.
A basic performance comparison to an existing modeling approach is given in \secref{sec:performance}.

\section{Modeling a Process Execution}
\label{sec:math-model}

This paper introduces BottleMod, a way of modeling the execution of a scientific workflow's task as a so-called \emph{process} in a generic mathematical fashion. A \emph{process} can also be used to model other events. An example is a data transfer over the network, as demonstrated later in \secref{sec:eval}.

Predicting the behavior of a task needs process-specific and execution-environment-specific knowledge. While a developer might know how a specific program works, the amount of resources available for the execution is typically unknown during development. On the other hand, the execution environment or the corresponding system administrator might have detailed knowledge about the available resources and how they will be allocated to the tasks. However, they do not know how the tasks work.
BottleMod is designed to distinguish between the process-specific requirements and execution-specific resource allocation.
This separation enables different parties to describe the requirements and resource allocation, resolving the issue. 
The information for the task model could either be statically annotated, measured, or learned from other executions of the same task. In the long run, BottleMod could be used for modeling, simulation, and analysis, providing valuable information for resource allocation and helping a dynamic scheduler with short- to midterm decisions. Also, the result visualization could be helpful in understanding how bottlenecks in an early stage can influence the execution later.

We model processes by their \emph{data} and \emph{resource} demands. Here, each process can have an arbitrary number of data and resource requirements.

\begin{enumerate}
  \item \emph{Data} is the input data directly available to the process. Input data can be stored for later usage. The model assumes that an unlimited amount of data can be stored, and that stored data does not expire.
  \item \emph{Resources} cannot be stored. Unused resources will not yield an advantage for the execution but are just gone. Therefore, resources help modeling the CPU time an application needs or the data rate of a link used for a data transfer.
\end{enumerate}

These requirements are defined by \emph{requirement} functions for each resource and input data per process. The functions describe how much of which input data or resource the process will need to progress until a certain point, measured by an abstract \emph{progress} metric. Accordingly, these functions are always monotonically increasing.
While requirement functions describe the requirements of a process, so-called \emph{input} functions for a process define the amount of available data input and available resource at each point in time. Notice that contrary to the data input, the allocated amount of a resource can be lowered during execution.

These functions contain all the necessary information to derive a \emph{progress} function, describing the progress of the process over time---representing its execution. \emph{Output} functions map the process's progress to the amount of data generated, enabling to calculate at which point in time how much usable output data is generated.
The dimensions for all these functions can be chosen freely, given they are consistent for those interacting.


\subsection{Progress Metric}

The progress metric $p$ defines how far the process has progressed. This metric does not need to relate to any resource or the estimated time to complete the process. It is completely arbitrary and chosen by the developer of the model. The progress metric must only be consistent inside a single process, i.e., all process-describing functions for the same process must use the same progress metric.

\subsection{Requirement Functions}

A requirement ($\mathcal{R}$) function describes how much input a process needs to achieve a certain progress. This relation corresponds to the requirements of a process independent of its execution environment. Typically, defining the process's needs would be characterized by the amount of data or resources needed at a specific time. However, such a direct definition would couple a process's own attributes with execution-environment-specific ones---including the availability of data and the resource allocation. Requirement functions are time-independent but only provide the input-to-progress relation to separate the process's requirements and allocated resources. Therefore, our approach is quite unconventional.

\subsubsection{Data Requirement Functions}


A \emph{data} requirement function defines the data requirement of a process. For each $k \in \{1,\ldots,K\}$, where $K$ is the number of relevant data inputs, the mathematical function $\mathcal{R}_{Dk}(n_{Dk})$ maps the amount of data input needed $n_{Dk}$ (in bytes, for example) to the maximum possible progress the process can make given there are no other limitations.
Those functions must be monotonically increasing since the possible progress cannot decrease when more input data becomes available.

\newcommand{\graphcoord}[2]{($ (\p1)!#1!(\x2,\y1) + (0,0)!#2!(0,\y2-\y1) $)}

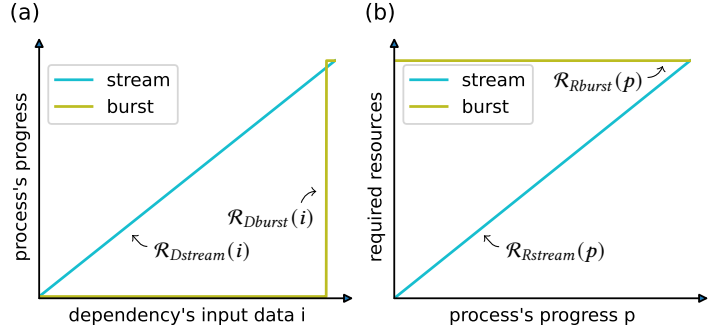
\begin{figure}[t!]
  \centering%
  \begin{tikzpicture}[textnode/.style={text height=1.5ex,text depth=.25ex}]
    \node[anchor=south west] (fig) at (0,0)
       {\includegraphics[width=.635\linewidth]{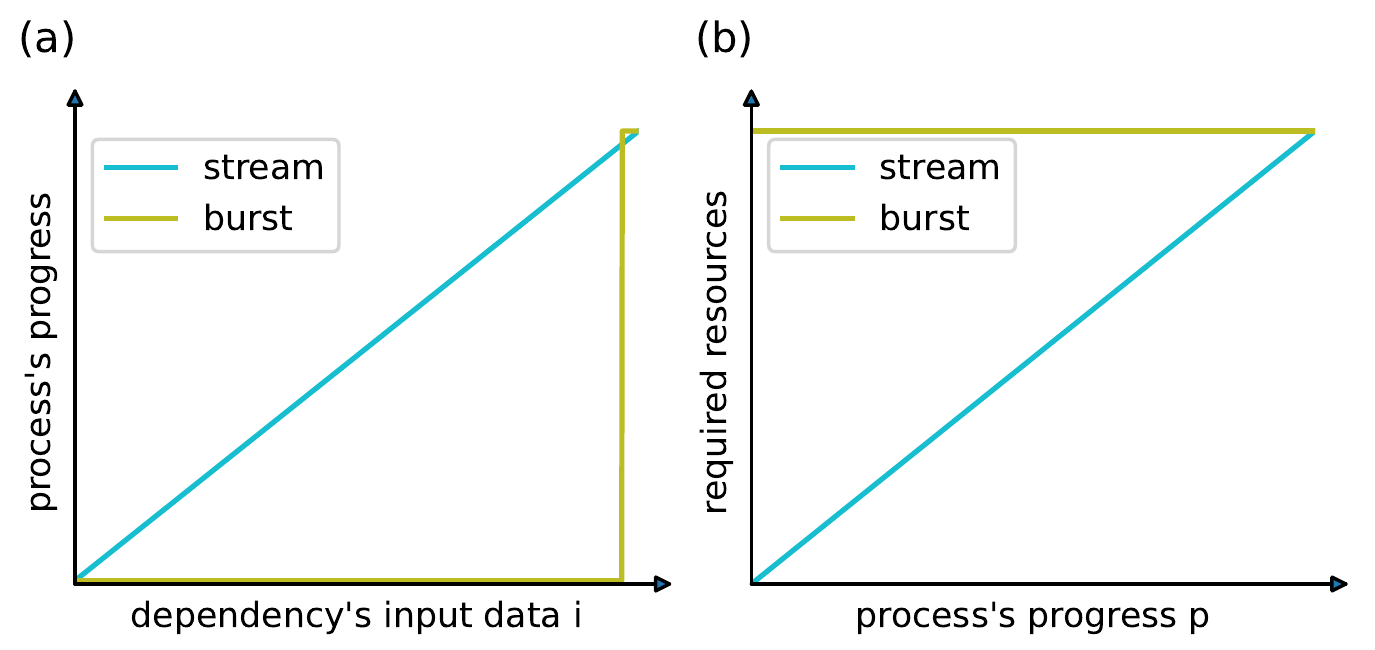}};
  \coordinate (originx) at ($ (fig.south west)!.07!(fig.south east) $);
  \coordinate (originy) at ($ (fig.south west)!.11!(fig.north west) $);
  \coordinate (maxx) at ($ (fig.south west)!.98!(fig.south east) $);
  \coordinate (maxy) at ($ (fig.south west)!.84!(fig.north west) $);

  \coordinate (aorigin) at ($ (originx) + (originy)!.03!(maxy) $);
  \coordinate (amax)    at ($ (originx)!.47!(maxx) + (originy)!1!(maxy) $);
  \coordinate (borigin) at ($ (originx)!.52!(maxx) + (originy)!.03!(maxy) $);
  \coordinate (bmax)    at (maxx |- maxy);

  \path let \p1 = (aorigin), \p2 = (amax) in
   node[inner sep=2pt,anchor=north west] at \graphcoord{.33}{.23} (od1) {\small\emph{$\mathcal{R}_{\mathit{Dstream}}(i)$}};
  \draw[->,>=to] (od1.175) to[out=180,in=-45] +(-2.mm,1.5mm);
  \path let \p1 = (aorigin), \p2 = (amax) in
  node[inner sep=2pt,anchor=east] at \graphcoord{.88}{.3} (od2) {\small\emph{$\mathcal{R}_{\mathit{Dburst}}(i)$}} (amax);
  \draw[->,>=to] (od2.27) to[out=70,in=180] +(2.mm,1.5mm);

  \path let \p1 = (borigin), \p2 = (bmax) in
  node[inner sep=2pt,anchor=north west] at \graphcoord{.33}{.23} (od1)
  {\small\emph{$\mathcal{R}_{\mathit{Rstream}}(p)$}};
  \draw[->,>=to] (od1.175) to[out=180,in=-45] +(-2mm,1.5mm);
  \path let \p1 = (borigin), \p2 = (bmax) in
   node[inner sep=1pt,anchor=north east] at \graphcoord{.79}{.89} (od2) {\small\emph{$\mathcal{R}_{\mathit{Rburst}}(p)$}};
  \draw[->,>=to] (od2.5) to[out=0,in=250] +(2.mm,1.5mm);
  \end{tikzpicture}
  \caption{Exemplary requirement functions.
  Part (a): common \emph{data} requirement functions, where `stream' is a data requirement function modeling a process that can progress with every new byte of the data dependency's input $i$ read (akin to stream processing). In contrast, `burst' is a data input where all data must first be read entirely by the process before progress can be made.
  Part (b): common \emph{resource} requirement functions. Here, `stream' shows a resource needed continuously to let the process make progress $p$. `Burst' shows a resource only required at the beginning, e.g., for a process that needs all its CPU time before it progresses at all.
  }
  \label{fig:samplerequirements}
\end{figure}


Two examples of such functions $\mathcal{R}_{Dk}(n_{Dk})$ are shown in \figref{fig:samplerequirements}\,(a). Simply reencoding a video would need just a little bit of input data for the process to start making progress. This would correlate to the function type displayed by `stream'. Reversing a video needs all input data before any progress can be made and would, therefore, correspond to `burst'.

\subsubsection{Resource Requirement Functions}

\emph{Resource} requirement functions work similarly to data requirement functions. For each resource $\ell \in \{1,\ldots,L\}$, where $L$ is the number of relevant resources, the mathematical function $\mathcal{R}_{R\ell}(p)$ yields the amount of resource needed (in a resource-dependent unit such as CPU cycles, transferred bytes, \ldots) at the progress $p$ of the process.


While the data requirement functions limit the maximum progress of a process, resources may only limit the progression speed. When using CPU time as an example, a resource function would define how much CPU time (y-axis) is needed to progress to a particular execution point (x-axis). I.e., if the process requires the same amount of CPU time to progress, e.g., reencoding a video, the corresponding resource requirement function would be a linear function with a constant slope (as seen as `stream' in \figref{fig:samplerequirements}\,(b)). Another example would be a process that needs all the CPU time before it can progress. Such an example is displayed as `burst' in \figref{fig:samplerequirements}\,(b).

\subsection{Input Functions}

Requirement functions alone do not suffice to model a process's actual execution. Additionally, other functions are needed for both the $K$ required data inputs and the $L$ relevant resources a process uses. The data input function $I_{Dk}(t)$ and the resource input function $I_{R\ell}(t)$ relate the real-time $t$ to the available amount of input data and resources for any point in time.

These functions are not directly related to the process itself but describe a particular execution environment, e.g., CPU time assigned by a scheduler, input data the process depends on from a network link, or a previous process.

Similar to the data requirement functions, the data input functions $I_{Dk}(t)$ describe how much input data is available at which point in time. They are monotonically increasing as input data provided once are not lost since all data are storable.

Resource input functions $I_{R\ell}(t)$, on the other hand, are neither accumulative nor necessarily monotonically increasing. They denote the resource amount per time unit available at the time $t$. For example, if a process is assigned the same amount of a CPU resource over time, the corresponding resource allocation function $i_{R\ell} = I_{R\ell}(t)$ would be a constant function. It would denote that a fixed number $i_{R\ell}$ of CPU cycles per time unit is allocated to the process's execution at every point in time.

\subsection{Data Output Functions}
\label{sec:output_functions}

The above functions already contain enough information to determine the execution behavior of the process, including execution time, resource, and data usage. Deriving this information is possible by calculating the progress function, which defines the progress made over time and is described in \secref{sec:process-sim}. This still does not define how much data the process generates. To conclude this, data output functions are needed.

Data output functions $o_{m} = O_{m}(p)$ define for each $m \in \{1,\ldots,M\}$, where $M$ is the number of data outputs, how much data $o_{m}$ is already generated (y-axis) at a certain progress $p$ (x-axis). They resemble inverted data requirement functions. In combination with the calculated progress function of a process, data output functions are used to determine how much data is generated at which point in time. This output is particularly useful since it is identical to data input functions and, therefore, can be used as an input in a successor process, which is explained in more detail in \secref{sec:comb}.

\section{Deducing the Behavior of a Process}
\label{sec:process-sim}

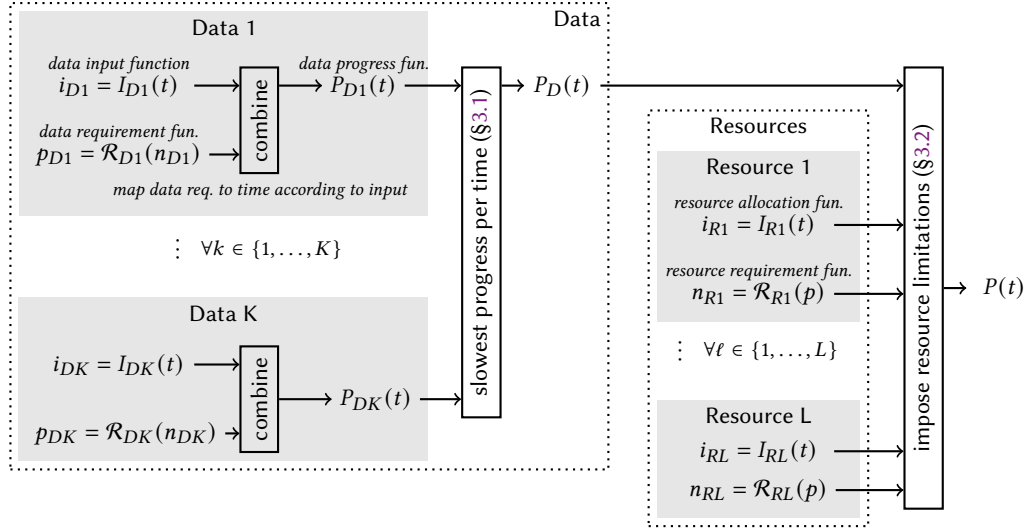
\begin{figure*}
    \begin{tikzpicture}[
      operationnode/.style={rectangle, draw=black, thick, minimum size=5mm},
      functionnode/.style={rectangle, rounded corners=3pt, minimum width=2.cm},
      groupnode/.style={dotted, draw=black, thick},
      groupnodeinner/.style={fill=white!90!black},
    ]
  \sffamily
    \node[functionnode]         (D1R)       []                         { $i_{D1} = I_{D1}(t)$ };
    \node[]                     (D1Iinfo)   [above=-2mm of D1R] {\rmfamily\footnotesize\emph{data input function}  };
    \node[functionnode]         (D1I)       [below=4mm of D1R] { $p_{D1} = \mathcal{R}_{D1}(n_{D1})$  };
    \node[]                     (D1Iinfo)   [above=-2mm of D1I] { \rmfamily\footnotesize\emph{data requirement fun.}  };
    \node[]                     (D1H)       [above=4mm of D1R]                   {};
    \path let \p1=(D1R.north), \p2=(D1I.south) in
      node[operationnode]       (D1O)
      [minimum height={\y1-\y2-\pgflinewidth}, right=0.6cm of D1R.north east,anchor=north west]   {\rotatebox{90}{combine}};
    \node[]                     (D1Oinfo)   [below=0mm of D1O] { \rmfamily\footnotesize\emph{~~map data req. to time according to input~~}  };
    \node[functionnode,minimum width=1cm]         (D1P)       [right=17mm of D1R]         { $P_{D1}(t)$ };
    \node[]                     (D1Pinfo)   [above=-2mm of D1P] { \rmfamily\footnotesize\emph{data progress fun.}  };

    \draw[->, thick] (D1R.east) -- (D1R.east -| D1O.west);
    \draw[->, thick] (D1I.east) -- (D1I.east -| D1O.west);
    \draw[->, thick] (D1O.east |- D1R.east) -- (D1P.west);

    \begin{scope}[on background layer]
    \node[groupnodeinner]            (D1)        [fit=(D1R)(D1I)(D1H)(D1O)(D1P)(D1Oinfo)]         {};
    \end{scope}
    \node[]                     (D1Label)   [anchor=north]  at (D1.north)   { Data 1 };

    \node[functionnode]         (D2)       [below=7mm of D1O]      { \rotatebox{90}{\ldots}\raisebox{2pt}{\small\quad$\forall k\in \{1,\ldots,K\}$} };

    \node[functionnode]         (D3R)       [below=32mm of D1R]      { $i_{DK} = I_{DK}(t)$ };
    \node[functionnode]         (D3I)       [below=4mm of D3R,xshift=1mm]            { $p_{DK} = \mathcal{R}_{DK}(n_{DK})$ };
    \node[]                     (D3H)       [above=3mm of D3R]                   {};
    \path let \p1=(D3R.north), \p2=(D3I.south) in
      node[operationnode]       (D3O)
      [minimum height={\y1-\y2-\pgflinewidth}, right=0.6cm of D3R.north east,anchor=north west]   {\rotatebox{90}{combine}};
    \node[functionnode,minimum width=1cm]         (D3P)       [right=7mm of D3O]         { $P_{DK}(t)$ };

    \draw[->, thick] (D3R.east) -- (D3R.east -| D3O.west);
    \draw[->, thick] (D3I.east) -- (D3I.east -| D3O.west);
    \draw[->, thick] (D3O.east) -- (D3P.west);

    \begin{scope}[on background layer]
    \node[groupnodeinner]            (D3)        [fit=(D3R)(D3I)(D3H)(D3O)(D3P)(D3P-|D1Oinfo.east)]         {};
    \end{scope}
    \node[]                     (D3Label)   [anchor=north]  at (D3.north)   { Data K };

    \path let \p1=(D1P.north), \p2=(D3P.south) in
      node[operationnode]       (DO)
      [minimum height={\y1-\y2-\pgflinewidth}, right=0.75cm of D1P.north east,anchor=north west]   {\rotatebox{90}{slowest progress per time (§\ref{sec:data-progress})}};
    \node[functionnode,minimum width=1cm]         (DP)        [right=0.3 cm of DO.east |- D1P]            { $P_D(t)$ };

    \draw[->, thick] (D1P.east) -- (D1P.east -| DO.west);
    \draw[->, thick] (D3P.east) -- (D3P.east -| DO.west);
    \draw[->, thick] (DO.east |- DP.west) -- (DP.west);

    \begin{scope}[on background layer]
      \node[groupnode]            (D)        [fit=(D1)(D2)(D3)(DO)(DP)]         {};
      \node[]                     (DLabel)   [anchor=north east]  at (D.north east)   { Data };
    \end{scope}

    \node[functionnode]         (R1R)      [right=16mm of DP |- D2,yshift=3mm] { $i_{R1} = I_{R1}(t)$ };
    \node[]                     (R1Rinfo)  [above=-2mm of R1R] { \rmfamily\footnotesize\emph{resource allocation fun.}  };
    \node[functionnode]         (R1I)      [below=4mm of R1R]       { $n_{R1} = \mathcal{R}_{R1}(p)$ };
    \node[]                     (R1Iinfo)  [above=-2mm of R1I] { \rmfamily\footnotesize\emph{resource requirement fun.}  };
    \node[]                     (R1H)      [above=4mm of R1R]       {};
    \begin{scope}[on background layer]
    \node[groupnodeinner]            (R1)       [fit=(R1R)(R1I)(R1H)(R1Rinfo)]      {};
    \node[]                     (R1Label)  [anchor=north]  at (R1.north) { Resource 1 };
    \end{scope}

    \node[functionnode]         (Rx)       [below=2mm of R1I]      { \rotatebox{90}{\ldots}\raisebox{2pt}{\small\quad$\forall\ell\in\{1,\ldots,L\}$} };

    \node[functionnode]         (R2R)      [below=25mm of R1R]       { $i_{RL} = I_{RL}(t)$ };
    \node[functionnode]         (R2I)      [below=0mm of R2R]       { $n_{RL} = \mathcal{R}_{RL}(p)$ };
    \node[]                     (R2H)      [above=1mm of R2R]       {};
    \begin{scope}[on background layer]
    \node[groupnodeinner]       (R2)       [fit=(R2R)(R2I)(R2H)(R2R-|R1Rinfo.west)(R2R-|R1Rinfo.east)]      {};
    \node[]                     (R2Label)  [anchor=north]  at (R2.north) { Resource L };
    \end{scope}

    \node[]                     (RH)       [above=2mm of R1]           {};
    \begin{scope}[on behind layer]
      \node[groupnode]            (R)        [fit=(R1)(R2)(RH)]            {};
    \end{scope}
    \node[]                     (RLabel)   [anchor=north]  at (R.north)  { Resources };

    \path let \p1=(DP.north), \p2=(R2I.south) in
      node[operationnode]       (O)
      [minimum height={\y1-\y2-\pgflinewidth}, right=9mm of R2I.south east,anchor=south west]   {\rotatebox{90}{impose resource limitations (§\ref{sec:resource-limits})}};
    \node[functionnode,minimum width=1cm]         (P)        [right=0.3cm of O]            { $P(t)$ };

    \draw[->, thick] (DP.east) -- (DP.east -| O.west);
    \draw[->, thick] (R1R.east) -- (R1R.east -| O.west);
    \draw[->, thick] (R1I.east) -- (R1I.east -| O.west);
    \draw[->, thick] (R2R.east) -- (R2R.east -| O.west);
    \draw[->, thick] (R2I.east) -- (R2I.east -| O.west);
    \draw[->, thick] (O.east |- P.west) -- (P.west);
  \end{tikzpicture}
  \caption{Overview of the steps required to calculate the progress function for a process. The data input functions $I_{Dk}(t)$ and data requirement functions $\mathcal{R}_{Dk}(n_{Dk})$ are combined to calculate the maximum possible progress $P_{Dk}(t)$ regarding a certain data input $k \in \{1,\ldots,K\}$. Taking the minimum of this progress over all data inputs yields the maximum possible progress data-wise $P_D(t)$. Together with all resource allocation functions $I_{R\ell}(t)$ and all resource requirement functions $\mathcal{R}_{R\ell}(p)$ ($\ell \in \{1,\ldots,L\}$) we can calculate the overall progress function $P(t)$. The iterative algorithm for the calculation is described in \secref{sec:resource-limits}.}
  \label{fig:flowchart}
\end{figure*}

Combining the process-specific requirement functions and execution-specific input functions allows for calculating the time a process needs to achieve progress. The amount progress a process makes until a point in time $t$ is represented by the \emph{progress} function $P(t)$. In this model, a process finishes when the maximum progress is reached. Analyzing a process's execution corresponds to calculating the progress function of that process.

Calculating the progress function happens in two steps (\figref{fig:flowchart}). First, the data limitations are combined to deduce the maximum progress possible regarding all process data inputs. Then, the effects of resource limits are taken into account to infer the process's overall progress.

\subsection{Data Progress}
\label{sec:data-progress}

For every pair of a data requirement function $\mathcal{R}_{Dk}(n_{Dk})$ and its respective data input function $I_{Dk}(t)$, we calculate a corresponding data progress function $P_{Dk}(t)$. This function defines the maximum progress limited by the specific data input $k$ over time.
Since $\mathcal{R}_{Dk}(n_{Dk})$ specifies the possible progress with a given amount of input data, and $I_{Dk}(t)$ defines the available input data at a given time, $P_{Dk}(t)$ can be calculated as follows:
\begin{equation}
  P_{Dk}(t) = \mathcal{R}_{Dk}(I_{Dk}(t)),\quad\forall k \in \{1,\ldots,K\}
\end{equation}

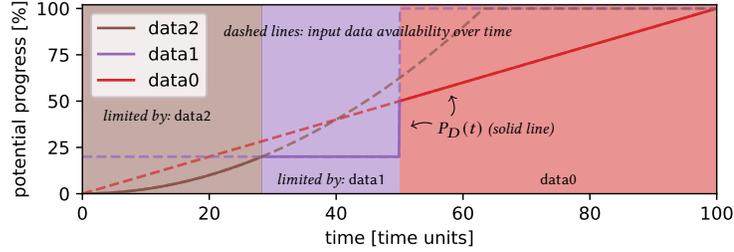
\begin{figure}
  \centering
  \begin{tikzpicture}[textnode/.style={text height=1.5ex,text depth=.25ex}]
    \node[anchor=south west] (fig) at (0,0)
         {\includegraphics[width=.635\linewidth]{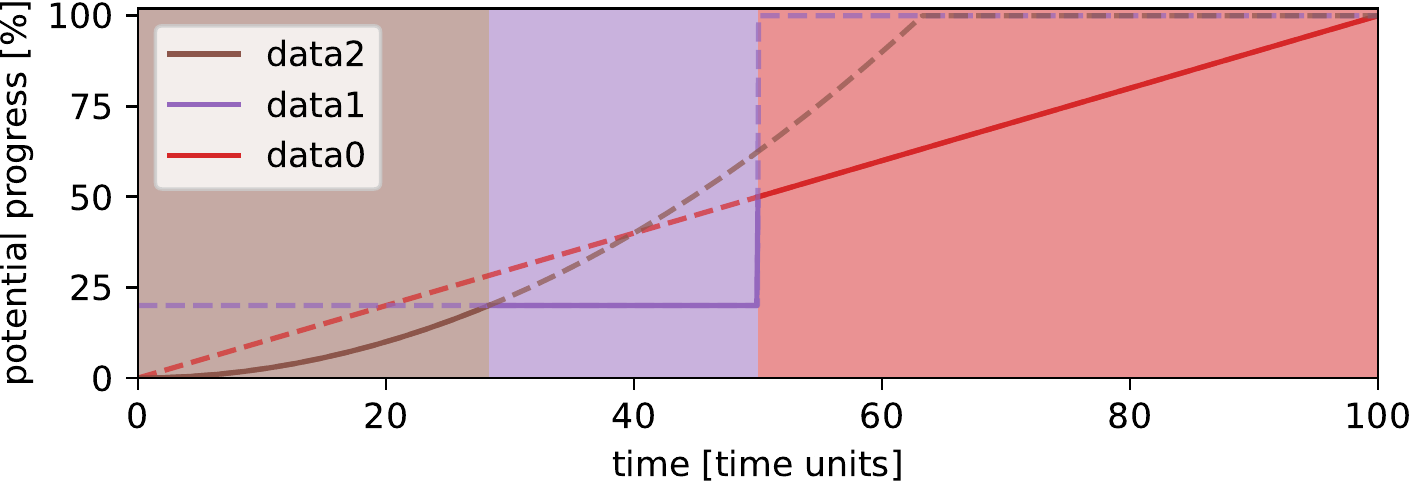}};
         \coordinate (originx) at ($ (fig.south west)!.11!(fig.south east) $);
         \coordinate (originy) at ($ (fig.south west)!.24!(fig.north west) $);
         \coordinate (maxx) at ($ (fig.south west)!.965!(fig.south east) $);
         \coordinate (maxy) at ($ (fig.south west)!.95!(fig.north west) $);
    \coordinate (progressorigin) at ($ (originx) + (originy) $);
    \coordinate (progressmax)    at ($ (maxx) + (maxy) $);
  \path let \p1 = (progressorigin), \p2 = (progressmax) in
  node[inner sep=2pt,anchor=north west] at \graphcoord{.55}{.41} (Pt) {\scriptsize\emph{$P_D(t)$ (solid line)}};
  \draw[->,>=to] (Pt.175) to[out=160,in=20] +(-2.8mm,-.3mm);
  \draw[->,>=to] (Pt.163) to[out=45,in=-70] +(0mm,2.7mm);
  \path let \p1 = (progressorigin), \p2 = (progressmax) in
  node[textnode,anchor=west] at \graphcoord{.015}{.42} {\scriptsize\emph{limited by:} \textsf{data2}};
  \path let \p1 = (progressorigin), \p2 = (progressmax) in
  node[textnode,anchor=south,inner sep=2pt] at \graphcoord{.39}{.001} {\scriptsize\emph{limited by:} \textsf{data1}}
  node[textnode,anchor=south,inner sep=2pt] at \graphcoord{.75}{.001} {\scriptsize\textsf{data0}};
  \path let \p1 = (progressorigin), \p2 = (progressmax) in
  node[textnode,inner sep=2pt,anchor=west] at \graphcoord{.21}{.87} {\scriptsize\emph{dashed lines: input data availability over time}};
  \end{tikzpicture}
  \caption{Example of data progress functions. Based on the data output and data input functions, three data progress functions are shown. `Data0'  increases linearly, `data1' starts with 20\,\% of available input and later jumps to 100\,\%, and `data2' increases quadratic over time. The dotted lines show the maximum possible progress $P_{Dk}(t)$ for the specific data input $k$, while the solid line shows $P_D(t)$, the maximal overall possible data progress. The background colors correspond to the data input that is currently the lowest and clearly indicate which data input keeps the process from making more progress.}
  \label{fig:dataprogress}
\end{figure}

These individual data progress functions (see \figref{fig:dataprogress} for an example) are combined into a function $P_D(t)$ that defines the maximum possible progress based on all data inputs. Since the actual progress can never be higher than the lowest of the data progress functions, $P_D(t)$ can be generated by section-wise choosing the function that yields the lowest value (solid line in \figref{fig:dataprogress}). We denote this selection as $\min$ in:
\begin{equation}
P_D(t) = \min(P_{D1}(t), \ldots, P_{DK}(t))
\label{eq:p_d}
\end{equation}
The overall progress $P(t)$ is not only limited by the data progress but also by the resources. Therefore, the data progress imposes an upper limit on the overall progress:
\begin{equation}
P(t) \le P_D(t) \label{eq:progress-datalimit}
\end{equation}

\subsection{Impose Resource Limitations}
\label{sec:resource-limits}

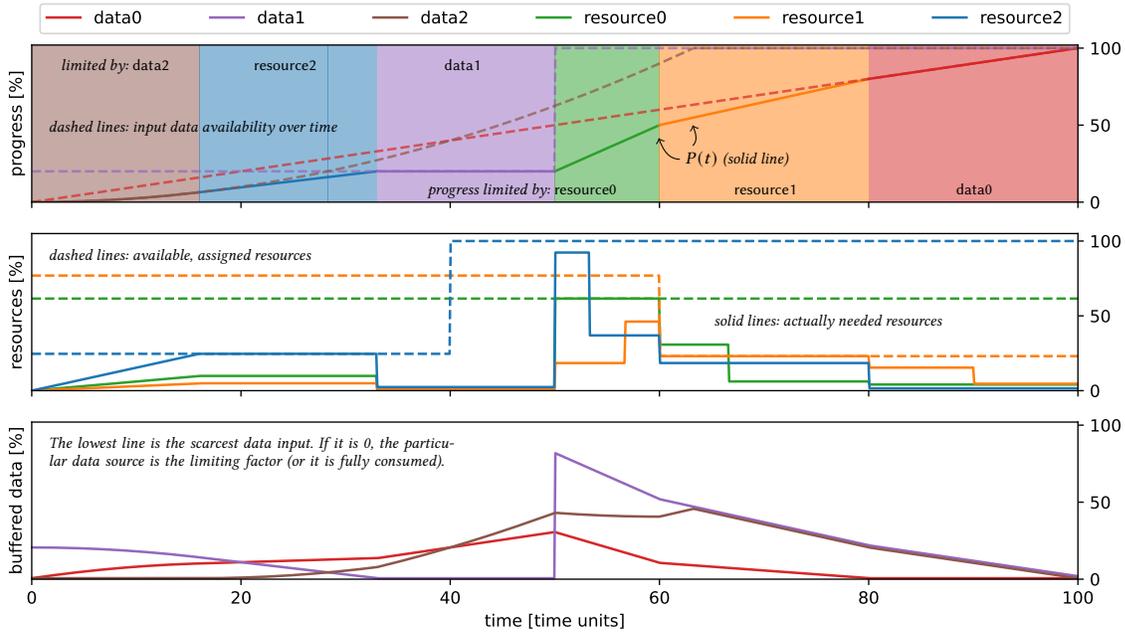
\begin{figure*}
  \centering
\begin{tikzpicture}[textnode/.style={text height=1.5ex,text depth=.25ex}]
  \node[anchor=south west] (fig) at (0,0)
       {\includegraphics[width=\linewidth]{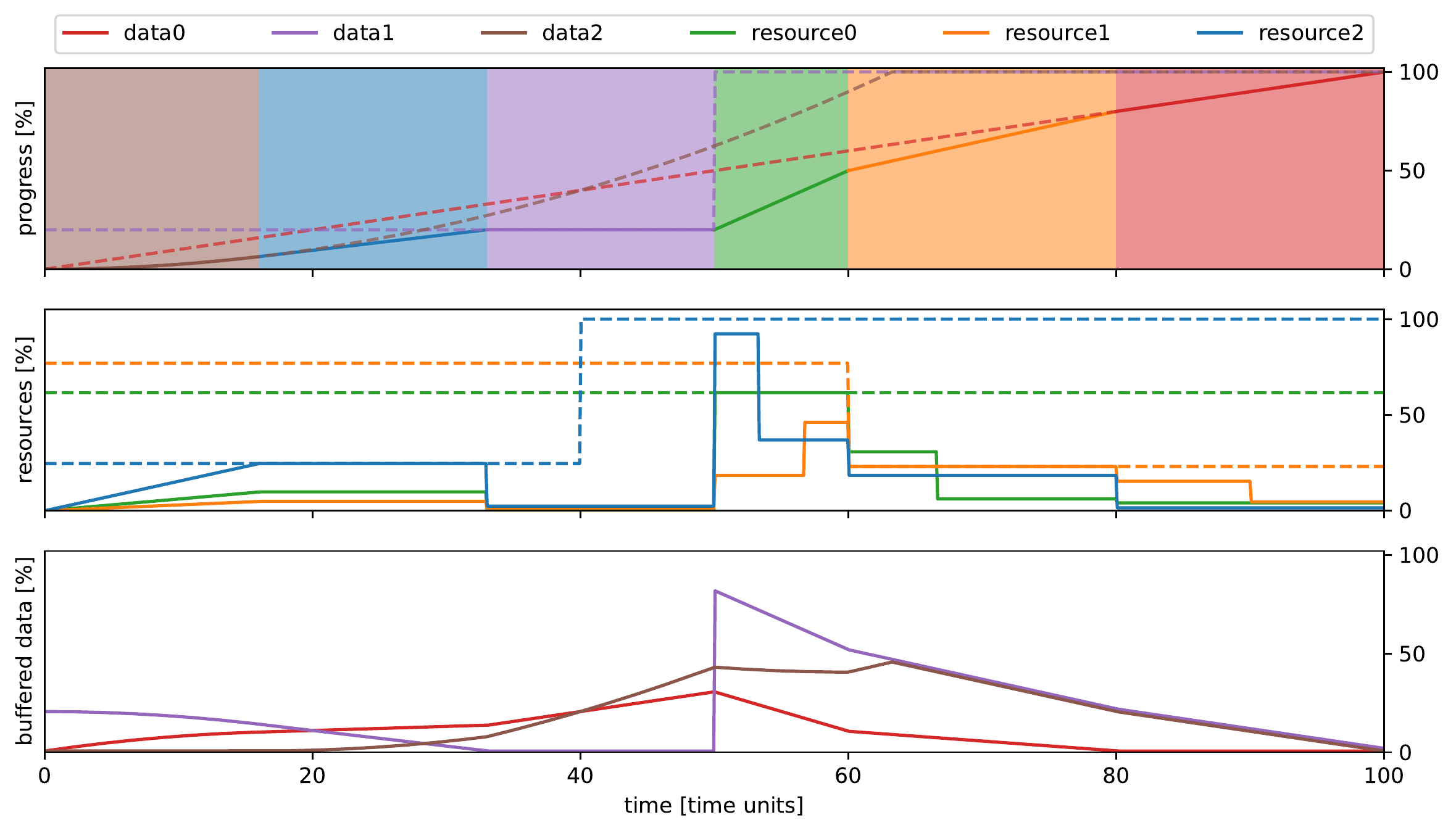}};
  \coordinate (originx) at ($ (fig.south west)!.04!(fig.south east) $);
  \coordinate (originy) at ($ (fig.south west)!.11!(fig.north west) $);
  \coordinate (maxx) at ($ (fig.south west)!.945!(fig.south east) $);
  \coordinate (maxy) at ($ (fig.south west)!.906!(fig.north west) $);
  \coordinate (progressorigin)  at ($ (originx) + (originy)!.7!(maxy) $);
  \coordinate (progressmax)     at (maxx |- maxy);
  \coordinate (resourcesorigin) at ($ (originx) + (originy)!.33!(maxy) $);
  \coordinate (resourcesmax)    at ($ (maxx) + (originy)!.64!(maxy) $);
  \coordinate (bufferedorigin)  at (originx |- originy);
  \coordinate (bufferedmax)     at ($ (maxx) + (originy)!.3!(maxy) $);
1  

   \draw let \p1 = (progressorigin), \p2 = (progressmax) in
   node[inner sep=2pt] at \graphcoord{.673}{.29}
     (Pt) {\scriptsize\emph{$P(t)$ (solid line)}} (progressmax);
  \draw[->,>=to] (Pt.west) to[out=180,in=-70] +(-2.8mm,2.8mm);
  \draw[->,>=to] (Pt.163) to[out=45,in=-70] +(0mm,2.7mm);
  
  \path (progressorigin |- progressmax) --
  node[textnode,pos=.077,below] {\scriptsize\emph{limited by:} \textsf{data2}}
  node[textnode,pos=.24,below] {\scriptsize\textsf{resource2}}
  node[textnode,pos=.41,below] {\scriptsize\textsf{data1}}
  (progressmax);
  \path (progressorigin) --
  node[textnode,pos=.467,above] {\scriptsize\emph{progress limited by:} \textsf{resource0}}
  node[textnode,pos=.70,above] {\scriptsize\textsf{resource1}}
  node[textnode,pos=.90,above] {\scriptsize\textsf{data0}}
  (progressmax |- progressorigin);
  \path let \p1 = (progressorigin), \p2 = (progressmax) in
  node[textnode,inner sep=2pt,anchor=west] at \graphcoord{.008}{.5}
   {\scriptsize\emph{dashed lines: input data availability over time}}
  (progressmax);

  \path let \p1 = (resourcesorigin), \p2 = (resourcesmax) in
  node[textnode,inner sep=2pt,anchor=west] at \graphcoord{.008}{.9}
  {\scriptsize\emph{dashed lines: available, assigned resources}}
  node[textnode,inner sep=2pt] at \graphcoord{.76}{.5}
  {\scriptsize\emph{solid lines: actually needed resources}};

  {\scriptsize
  \path let \p1 = (bufferedorigin), \p2 = (bufferedmax) in
  node[textnode,inner sep=2pt,anchor=west,text width=55mm] at \graphcoord{.008}{.84}
  {\scriptsize\emph{The lowest line is the scarcest data input. If it is 0, the particular data source is the limiting factor (or it is fully consumed).}}
  (bufferedmax);  node[textnode,pos=.68,below=18mm,inner sep=2pt,rotate=-6] {\scriptsize\emph{lowest line is scarcest data input}}
  (bufferedmax);
  }
\end{tikzpicture}
  \caption{
    Top: Final progress function $P(t)$ (solid) along with data-based progress functions $P_{Dk}(t)$ (dashed). The color of $P(t)$ and the background indicate the input data or resource that limits the progress in this section.
    Mid: Resource consumption $P'(t) \cdot \mathcal{R}_{R\ell}'(P(t))$ (solid) and assigned resources $I_{R\ell}(t)$ (dashed) for the three resources of the example.
    Bottom: Buffered input data from the process not yet used, $I_{Dk}(t) - \mathcal{R}_{Dk}^{-1}(P(t))$.
  }
  \label{fig:final}
\end{figure*}

Calculating the maximum possible progress considering resource demands is more complex than the deduction of data progress, as resources are not storable.

According to our model, the derivative of $\mathcal{R}_{R\ell}(p)$ defines for a process the amount of a resource $\ell$ needed to achieve the $p$-th progress, while $P(t)$ defines the achieved progress over time. In turn, $\mathcal{R}_{R\ell}'(P(t))$ results in a \emph{resource per progress} metric, describing the needed amount of a resource to generate a marginal amount of progress at time $t$. This expression defines how expensive progress is at time $t$ (e.g., how much CPU time per progress is needed for every additional progress unit). The required amount of each resource also depends on the progress speed, which is defined by the derivative $P'(t)$. The resource amount a process needs at time $t$ is thus:
\begin{equation}
  P'(t) \cdot \mathcal{R}'_{R\ell}(P(t))
\end{equation}

As long as this function does not exceed $I_{R\ell}(t)$, there is no limitation by the resource as the amount of available resources always surpasses or satisfies the demand. In such a case, only the data limitation would apply so that $P(t)$ would equal $P_D(t)$, which corresponds to the maximum possible data progress (\secref{sec:data-progress}).

When the amount of resources needed is greater than the amount provided, the progress has to slow down. To determine how much the progress has to be slowed down, the \emph{speedup} metric $S_{R\ell}(t)$ is used, which is given by:
\begin{equation}
  S_{R\ell}(t) = \frac{I_{R\ell}(t)}{P'(t) \cdot \mathcal{R}_{R\ell}'(P(t))}  \label{eq:resource-speedup}
\end{equation}
If the speedup is greater than $1$, the progress could be faster according to this resource. If the speedup is less than $1$, the progress is faster than the allocation of this resource allows. In such a case, $P(t)$ has to be adjusted by the `speedup' factor.

To impose such limitations, we first assume that $P(t)$ equals $P_D(t)$. Afterward, $S_{R\ell}(t)$ is calculated for each of the $L$ resources.
These speedup functions are combined into one function using the minimum. This combination follows the approach in equation \eqref{eq:p_d} for single data inputs. We then identify the first position where the combined speedup function drops below $1$ and call this position $t_x$. It is the first time a \emph{resource} limits progress. Therefore, the progress function is assumed to be correct up to $t_x$. The combined speedup function helps to determine the new progress function. It is updated as follows:
\begin{equation}
    P(t) = \min\Big(P_D(t),
    \int{P'(t) \cdot \min_{\forall \ell\in\{1,\ldots,L\}}{\big(S_{R\ell}(t)\big)}~dt}\Big)
  \label{eq:apply-resource-limit}
\end{equation}
The maximum progress still is limited by $P_D(t)$, and the maximum slope of the progress has to be adjusted by the minimum of all limiting speedup factors $S_{R\ell}(t)$.

Since all $S_{R\ell}(t)$ depend on $P(t)$ themselves as seen in \eqref{eq:resource-speedup}, calculating $P(t)$ is tantamount for solving a differential equation. Whether this is possible and how difficult it is to solve depends on the actual functions, especially $\mathcal{R}_{R\ell}(p)$.

Instead of solving the differential equation in one step, it is possible to use an iterative approach. The speedup factors are calculated based on the last $P(t)$ and applied to it as seen in \eqref{eq:apply-resource-limit}. After one iteration, $P_{\mathit{new}}(t)$ will not be limited at $t_x$ anymore. Since $P_{\mathit{new}}(t)$ might differ from $P(t)$ after $t_x$, a new resource limitation can occur.
In $P_{\mathit{new}}(t)$ certain positions after $t_x$ might be limited heavier than needed because the limitation itself is based on the former $P(t)$. $P_{\mathit{new}}(t)$ might not have a certain resource limitation where the former had one. That is compensated as the multiplication by the speedup factor will speed up the progress in such locations.

If $P_{\mathit{new}}(t)$ is now assumed to be the progress function and the above steps are repeated, $t_x$ will be greater than last time---guaranteeing progress in resolving the equation. This iterative algorithm, also shown in \algoref{algo:resourcelimit1}, can be used to change $P(t)$ until it is stable, i.e., it remains the same when going through the procedure described above. These arguments also imply that the combined minimum of the speedup factors is never smaller than $1$. An example of such a result can be seen in \figref{fig:final}.

\begin{algorithm}
  \caption{Impose Resource Limitations}
  \label{algo:resourcelimit1}

  \SetKwFunction{ppolymin}{min}
  \SetKwFunction{integrate}{Integrate}
  \SetFuncSty{text}

  $P_D \gets \ppolymin(P_{D1},\ldots,P_{DK})$\;
  $P \gets null$\;
  $P_{\mathit{new}} \gets P_D$\;

  \While{$P_{\mathit{new}} \neq P$}
  {
    $P \gets P_{\mathit{new}}$\;
    \For{$\ell$ in $1,\ldots,L$}
    {
      $S_{R\ell} \gets I_{R\ell}/(P_{\mathit{new}}' \cdot \mathcal{R}_{R\ell}'(P_{\mathit{new}}))$\;
    }
    $S_{lim} \gets \ppolymin(S_{R1}, \ldots, S_{RL})$\;
    $P_{\mathit{new}} \gets \ppolymin(P_D, \integrate(P_{\mathit{new}}' \cdot S_{lim}))$\;
  }
\end{algorithm}

The disadvantage of the iterative approach is that it may iterate over every $t$, which is not tractable in a practical implementation. Simplification of the algorithm by limiting the functions' complexity, as shown in \secref{sec:impl}, is a method to overcome that issue. However, the above algorithm (shown in \algoref{algo:resourcelimit1}) shows a possible way that theoretically works on any generic function type.

\subsection{Additional Simulation Information}

The final progress function $P(t)$ of a task predicts how a process will behave at a particular time. Besides allowing to derive when a process starts or finishes outputting its results through its output functions $O_{Dm}$ (\secref{sec:output_functions}), the progress function implicitly indicates the used fraction of a given resource and the amount of buffered input data.

\subsubsection{Resource Usage}

As described in \secref{sec:resource-limits}, the term $P'(t) \cdot \mathcal{R}_{R\ell}'(P(t))$ yields how much of a specific resource $\ell$ is needed at a particular time. For our example, the resource usage function is shown in \figref{fig:final}. Since the actual progress function $P(t)$ is known after the analysis, it is possible to use this term to calculate the resource consumption. The following term defines the relative resource usage for a specific resource $\ell$.
\begin{equation}
  \frac{P'(t) \cdot \mathcal{R}_{R\ell}'(P(t))}{I_{R\ell}(t)}  \label{eq:relative-resource-usage}
\end{equation}

This function expresses how much of the allocated amount of a resource is used. The resource usage will always be between $0$ and $1$. If the function is above $1$ at any point, it is either a rounding error or the implementation of the algorithm described in \secref{sec:resource-limits} is incorrect. Note that $I_{R\ell}(t)$ can be $0$ if nothing of the resource is allocated for the process at time $t$.

Whenever the resource usage equals the resource input function (relative resource usage of $1$ or resource allocation $I_{R\ell}(t)$ is $0$ while $\mathcal{R}_{R\ell}'(P(t)) \neq 0$), the resource is a bottleneck. Providing more of that resource (increasing $I_{R\ell}(t)$) \emph{could} speed up the progress if no other resource or data input is limiting the progress at the same time. Changing the input function for a point in time $t_y$ results in a change in $P(t)$ and, therefore, also may change the resource usage. In other words, changing any input function starting at $t_y$ might completely change the progress and the resource usage for all resources after $t_y$.

Similar to the relative resource usage \eqref{eq:relative-resource-usage}, other metrics can be calculated. Examples are the `\emph{amount of allocated but unused resource}' or `\emph{amount of resources needed so that only the data inputs impose a bottleneck}'.


\subsubsection{Buffered Data}

The evaluation of the data usage of a process works similarly. In this case, we need to find the inverse of the data output function. If this is possible, $\mathcal{R}_{Dk}^{-1}(P(t))$ defines how much data was consumed by the process until $t$. The amount of unused buffered data can be calculated using the following term:
\begin{equation}
  I_{Dk}(t)-\mathcal{R}_{Dk}^{-1}(P(t)) \label{eq:buffered-data}
\end{equation}

The term \eqref{eq:buffered-data} defines the amount of data provided in advance but not yet used by the process. \figref{fig:final} shows this metric for our example. Such a metric could help a scheduler to throttle the data input. This capabilty could be helpful when allocating the resources of a previous process that generates this input data, a previous process whose output is the data dependency in question.

\subsection{Combining Processes}
\label{sec:comb}

After calculating the progress function, the data output functions can be used to calculate how much data $o_m$ is generated until which point in time for a data output $m$:

$$ O_m(P(t)) $$

The resulting function has the same properties as a data input function. With this property, multiple processes can be chained by using this function as a data input function of a following process. That way, $O_m(P(t))$ of the process in the chain becomes $I_{Dk}(t)$ of the next process we investigate. A process can depend on several other processes, and the data output of one process can be an input function for multiple others. Complex workflows can be modeled using this approach. Cyclic dependencies are a potential limitation in the current model. As long as the dependency graph of all processes is acyclic, each process can be analyzed in sequential order on its own.

This modularity is beneficial when modeling a workflow with its data transfers. Since moving the data can be a severe bottleneck, taking this into account can be important for an accurate analysis. A process that models a network transfer is quite similar to a process that models the execution of a program. Such a process would typically have one data input representing the data that should be transmitted. Additionally, it would have one resource that models the data rate of the connection instead of, e.g., the CPU time. The input and the output could be connected to other processes. Like a scheduler allocates CPU usage and guarantees that the sum does not exceed the available CPU time, the input functions for transfer processes that share a network link would have to be managed accordingly. The chaining also allows the modeling of more sophisticated properties, such as the informed prioritization of data streams that share a network bottleneck. It also enables the modeling of many throughput-limited systems. Other examples are network topologies or the throughput of reading from a storage device.

\section{BottleMod Implementation}
\label{sec:impl}

When implementing the above model, a central question is how to represent and model the mathematical functions. Simple functions, e.g., constants, enable easier operations but limit the expressiveness to model task behavior. Supporting any function gives much freedom in modeling tasks and resource allocations but makes the implementation difficult and might result in slow analysis. To evaluate the ideas of this paper, the model presented above was implemented based on \emph{piecewise polynomials}, which define a function in several pieces, with each piece described as a polynomial. This approach enables the modeling of non-linear functions, hard edges, and jumps in functions. Of course, data functions and resource requirement functions must still be monotonically increasing across and in all its pieces.

Using an implementation for piecewise-defined polynomial functions such as PPoly from scipy~\cite{scipy} might induce problems. \secref{sec:resource-limits} has shown that the division operation on functions is required. The result of the division might not be a (piecewise-defined) polynomial function itself and may therefore not be representable in a framework such as PPoly. Even enabling negative exponents will not solve the problem since the derivative of these extended polynomials might result in logarithmic expressions. These could be used as a divisor again, making fully supporting polynomial functions here difficult and expensive.

One way to counteract the problem is to extend the representation of functions for logarithmic terms. We overcame this issue by restricting the resource requirement functions $\mathcal{R}_{R\ell}(p)$ to be piecewise-linear functions, which is the most typical way resources are provided in practice anyway. That also enables another way of solving the mathematical problem from \secref{sec:resource-limits}, which ensures that the speedup function $S_{R\ell}(t)$, defined in \eqref{eq:resource-speedup}, is always smaller than $1$ for every resource $\ell$. This equation can then be simplified as:
\begin{align}
  1 &\le \min_{\forall \ell\in\{1,\ldots,L\}}\left(\frac{I_{R\ell}(t)}{P'(t) \cdot \mathcal{R}_{R\ell}'(P(t))}\right) \nonumber \\
  P'(t) &\le \min_{\forall \ell\in\{1,\ldots,L\}}\left(\frac{I_{R\ell}(t)}{\mathcal{R}_{R\ell}'(P(t))}\right)
  \label{eq:easier-resource-limitations}
\end{align}

Since the resource requirement functions are piecewise-linear and only their derivative is used, the divisor in \eqref{eq:easier-resource-limitations} is piecewise-constant. Therefore, the result of the division is still representable by a piecewise polynomial. It also means that for a specific range in time, where the resource requirement function has a constant slope, the calculation of $P'(t)$ does not directly depend on the value of $P(t)$. That lets us easily calculate the maximum progress slope without solving a differential equation or iterating over every $t$. The algorithm iterates over the pieces of the resource requirement functions, calculates the maximum progress, updates $P(t)$ accordingly (which might change the later behavior of that function), and starts the next iteration. This approach is easier to implement than the generic solution presented in \secref{sec:resource-limits}. The adapted algorithm is shown in \algoref{algo:resourcelimit2}.

\begin{algorithm}
  \caption{Practically Impose Resource Limitations}
  \label{algo:resourcelimit2}

  \SetKw{kwOr}{or}
  \SetKw{kwAnd}{and}

  \SetKwFunction{ppolymin}{min}
  \SetKwFunction{integrate}{Integrate}
  \SetKwFunction{isgap}{IsGap}
  \SetKwFunction{nextlim}{NextLimitChg}
  \SetFuncSty{text}

  $P_D \gets \ppolymin(P_{D1},\ldots,P_{DK})$\;
  $P \gets P_D$\;
  $cur \gets 0$\;

  \While{$P(cur) < outputSize$}
  {
    $next \gets \infty$\;
    $maxSpeed \gets \ppolymin(\frac{I_{R1}}{\mathcal{R}_{R1}'(P)},\ldots,\frac{I_{RL}}{\mathcal{R}_{RL}'(P)})$\;
    \eIf{$P'(cur) > maxSpeed(cur)$ \kwOr $P'(cur) = maxSpeed(cur)$ \kwAnd $P''(cur) > 0$}
    {
      $P_{\mathit{new}} \gets \integrate(maxSpeed)$\;
      $maxSpeed \gets \ppolymin(\frac{I_{R1}}{\mathcal{R}_{R1}'(P_{\mathit{new}})},\ldots,\frac{I_{RL}}{\mathcal{R}_{RL}'(P_{\mathit{new}})})$\;
      $next \gets \nextlim(P, P_D, maxSpeed)$\;
      $P[cur \ldots next] \gets P_{\mathit{new}}$\;
    }{
      $next \gets \nextlim(P, P_D, maxSpeed)$\;
    }

    $cur \gets next$\;
  }
\end{algorithm}
Note that the pseudocode detects gaps in $P(t)$ by `NextLimitChg' and assumes that the slope at gaps is infinite. In a practical implementation with PPoly, such gaps with infinite slope are not easily recognized but must be detected by piece borders instead.
As iterations will often have a piece border as $cur$, concrete values like $P(cur)$ and $P'(cur)$ or also $maxSpeed(cur)$ are mathematically problematic. As the algorithm iterates with increasing $t$ the relevant value is the one using the piece that is also used with increasing $t$ (the piece on the `right' not the `left' one).

The performance of calculating the progress of a process depends not only on the amount of data and resource dependencies but also on the actual requirement and input functions of those dependencies. More sections in a piecewise function result in more iterations and operations. The complexity further depends on the number of changes of the limiting factor. The more often the limiting factor (a concrete input data or resource dependency) changes, the more iterations are needed.

Also, the order of the polynomials is critical since most operations become more complex when using higher-order polynomials, especially the operation of finding the roots.

For many typical situations, even piecewise-linear functions are sufficient. Finding the roots with them is simpler and can be done without precision loss because only rational numbers are needed. Another advantage is the possibility to invert (piecewise-defined) linear functions as long as their slope is non-zero.

We implemented the proposed model in Python. The piecewise-defined polynomial functions were implemented by using and extending the PPoly class from SciPy~\cite{scipy}.

\section{Evaluation}
\label{sec:eval}

To evaluate the proposed model, we modeled a small workflow and executed it under different circumstances. The measured execution times are then compared with the predictions of our BottleMod implementation.

\subsection{Example Workflow}

The example workflow is illustrated in \figref{fig:evalworkflow}. The three tasks of the workflow are single processes of the versatile video tool ffmpeg~\cite{ffmpeg} with each having a different job:

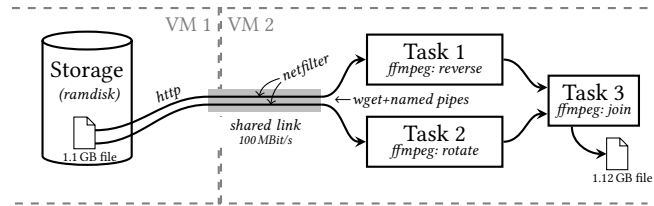
\begin{figure}
  \centering
  \begin{tikzpicture}[
  squarednode/.style={rectangle, draw=black, thick, minimum size=5mm, align=center},
  storagenode/.style={cylinder, shape border rotate=90, shape aspect=.2, draw=black, thick, yshift=-.5ex,
                      align=center},
  bottleneck/.style={rectangle, fill=black!25, minimum width=15mm, minimum height=3mm},
  labelml/.style={fill=white,align=center,text width=1.2cm,inner sep=1pt},
  label/.style={inner sep=0},>=stealth
]

  \coordinate           (Reference)                                                             {};
  \node[squarednode,minimum width=1.8cm]    (Task1)       [above=0.2cm of Reference]
       { Task 1 \\[-.2cm] \scriptsize \emph{ffmpeg: reverse}};
  \node[squarednode,minimum width=1.8cm]    (Task2)       [below=0.2cm of Reference]
       { Task 2 \\[-.2cm] \scriptsize \emph{ffmpeg: rotate}};
  \node[squarednode]    (Task3)       [right=1.5cm of Reference]
       { Task 3 \\[-.2cm] \scriptsize \emph{ffmpeg: join}};
  \node[bottleneck]     (Bottleneck)  [left=1.5cm of Reference] {};
  {\scriptsize
  \node[labelml]          (LabelBn)     [below=0.05cm of Bottleneck] { \emph{shared link\\[-2pt]\tiny{100\,MBit/s}} };
  \node[label,inner sep=1pt] (Netfilter) [anchor=south west,xshift=.6cm, above=0.15 of Bottleneck, rotate=22] { \emph{netfilter}};
  }

  \path let
  \p1=(Task1.north), \p2=(Task2.south), \n1 = {\y1-\y2-\pgflinewidth}
  in
  node[storagenode]   (Storage)  [left=1cm of Bottleneck, minimum height=\n1]
    { Storage \\[-.1cm]\mbox{\scriptsize{ \emph{(ramdisk)}}} \\[.3cm] \mbox{}};

    \draw (Storage.center) ++(-0,-.15)
    -- ++(-5pt,0) 
    -- ++(0,-12pt) 
    -- node[xshift=2pt,below=-2pt] {\tiny1.1\,GB file} ++(8pt,0) coordinate[yshift=2pt] (fileout1) 
    -- ++(0,9pt) coordinate[yshift=-2pt] (fileout2) 
    -- cycle 
    -- ++(0,-3pt) -- ++(3pt,0); 

  \node[anchor=west] at (Bottleneck.east) {\scriptsize\emph{~$\leftarrow$wget+named pipes}};

  \begin{scope}[on background layer]
  \draw[gray,dashed,thick] (Storage.north) ++(-1,0.3) -- ++(2.78,0) node[anchor=north east] {\small VM 1}
  -- ++(0,-2.6) -- ++(-2.78,0);

  \draw[gray,dashed,thick] (Storage.north -| Task3.east) ++(.1,0.3) -- ++(-5.7,0) node[anchor=north west] {\small VM 2}
  -- ++(0,-2.6) -- ++(5.7,0);
  \end{scope}

  \draw [->, thick] (Task1.east) to[out=0,in=180] ([yshift=5pt] Task3.west);
  \draw [->, thick] (Task2.east) to[out=0,in=180] ([yshift=-5pt]Task3.west);

  \draw [thick] (fileout2) to[out=0,in=180] node[sloped,pos=.7,above=-2.5pt] {\scriptsize\emph{http}} ([yshift=1.5pt] Bottleneck.west);
  \draw [thick] (fileout1) to[out=0,in=180] ([yshift=-1.5pt]Bottleneck.west);

  \draw [thick] ([yshift=1.5pt] Bottleneck.west) -- coordinate[pos=.45] (onlink1) ([yshift=1.5pt] Bottleneck.east);
  \draw [thick] ([yshift=-1.5pt] Bottleneck.west) -- coordinate[pos=.55] (onlink2) ([yshift=-1.5pt] Bottleneck.east);

  \draw[->,>=to] (Netfilter.west) to[out=195,in=80] (onlink1);
  \draw[->,>=to] (Netfilter.188) to[out=225,in=80] (onlink2);

  \draw [->, thick] ([yshift=1.5pt] Bottleneck.east) to[out=0,in=180] (Task1.west);
  \draw [->, thick] ([yshift=-1.5pt] Bottleneck.east) to[out=0,in=180] (Task2.west);

  \draw (Task3.315) ++(-0,-.15)
    -- ++(-5pt,0) 
    -- coordinate (result) ++(0,-12pt) 
    -- node[xshift=2pt,below=-2pt] {\tiny 1.12\,GB file} ++(8pt,0)  
    -- ++(0,9pt) coordinate[yshift=-2pt] (fileout2) 
    -- cycle 
    -- ++(0,-3pt) -- ++(3pt,0); 

    \draw [->, thick] (Task3.230) to[out=270,in=180] (result);

\end{tikzpicture}
  \caption{DAG of the demonstration workflow. The two inputs of task 1 and task 2 share a common bottleneck in form of a shared network link.}
  \label{fig:evalworkflow}
\end{figure}

\begin{itemize}
  \item Task 1 reverses the input video. It has to read and decode the total input video file before it can output the first byte of the reversed and encoded video. The encoded and reversed output video is much smaller than the input video due to encoding parameters. This task should represent a typical read-process-write process.
  \item Task 2 rotates the input video. That is just a change in the video container's metadata, and the task does not need to reencode the rotated video. As only metadata is changed and the actual video content is just copied, the task operates in a stream-processing way, writing the output concurrently to reading new input. Almost no CPU time is needed for that.
  \item Task 3 takes two input video files and writes their included video streams in a single output video file. As in task 2, the operation is done as stream processing without re-encoding and needs almost no CPU time.
\end{itemize}
The command lines for the three tasks used in this evaluation can be seen in the appendix \secref{sec:appendix}.

All tasks run on the same system. Although only task 1 needs a considerable amount of CPU time, the encoding part was limited to two cores by the command line option `\texttt{--threads 2}' to make sure that CPU time explicitly does not act as an additional shared bottleneck on the eight core system. System memory was also available in abundance to support the concurrent execution of all tasks.

If the input file is available locally and resides in a ramdisk, the execution of task 1 produces an 80\,MB video file in 108 seconds. Reading the input file and decoding it for later reversal takes 26 seconds. In the remaining 82 seconds, the result is encoded and written. The execution of task 2 takes 5 seconds and task 3 needs 3 seconds. Task 2 and 3 are limited purely by IO speed and need negligible CPU time.

Both task 1 and task 2 retrieve their input videos from the same remote webserver. The input for both task 1 and task 2 is the same video file, which resides in a ramdisk of the webserver. However, it is downloaded independently for both tasks. The video is a 5 minute long full HD clip with a total size of about 1.1\,GB (exactly 1,137,486,559\,bytes). Using the whole 100\,Mbit/s-link during workflow execution, a direct download of the video takes 89 seconds, meaning the net data rate is roughly 97.51\,Mbit/s.

The systems that run the workflow and the webserver are virtual machines virtualized with VMware on the same host connected to the same network. The host uses a Ryzen 2700X CPU, which had all cores fixed at 4.0\,GHz and its turbo disabled, and 48\,GB of main memory. Two cores with 4\,GB main memory were assigned to the webserver VM, and eight cores with 32\,GB were assigned to the workflow VM.

The bandwidth limitation of both connections accessing the webserver was achieved using netfilter~\cite{netfilter} via nftable's `limit rate' statement. The sum of both limits was 100\,Mbit/s. Since task 1 first completely downloads the video and processes it afterwards, the remote link was not directly used as input for ffmpeg. Instead, wget downloads the file to a named pipe with a buffer size set to 256\,MB, which is used as input for ffmpeg. That way, the network limit of task 2 can be adjusted to the whole 100\,Mbit/s when wget terminates. The overall execution time for the workflow differs depending on the starting limits of both connections.
The netfilter commands to handle the bandwidth limitation can be found in the appendix \secref{sec:appendix}.

Task 1 and 2 write their output data each into temporary files which serve as input for task 3. Task 3 is started after both task 1 and 2 are completed. Task 1 and 2 as well as the download with wget for task 1 are started at the same time.

This setup simulates a workflow with two typical ways of data consumption (represented by task~1 and task~2) and two potential bottlenecks: (i) the network link and (ii) the CPU resource of task~1. Only the network link limitation is a shared bottleneck. Therefore, no special scheduling had to be implemented and the evaluation remains comprehensible by only having one shared bottleneck.

\subsection{Model}

\begin{figure}
  \centering
  \includegraphics[width=.8\linewidth]{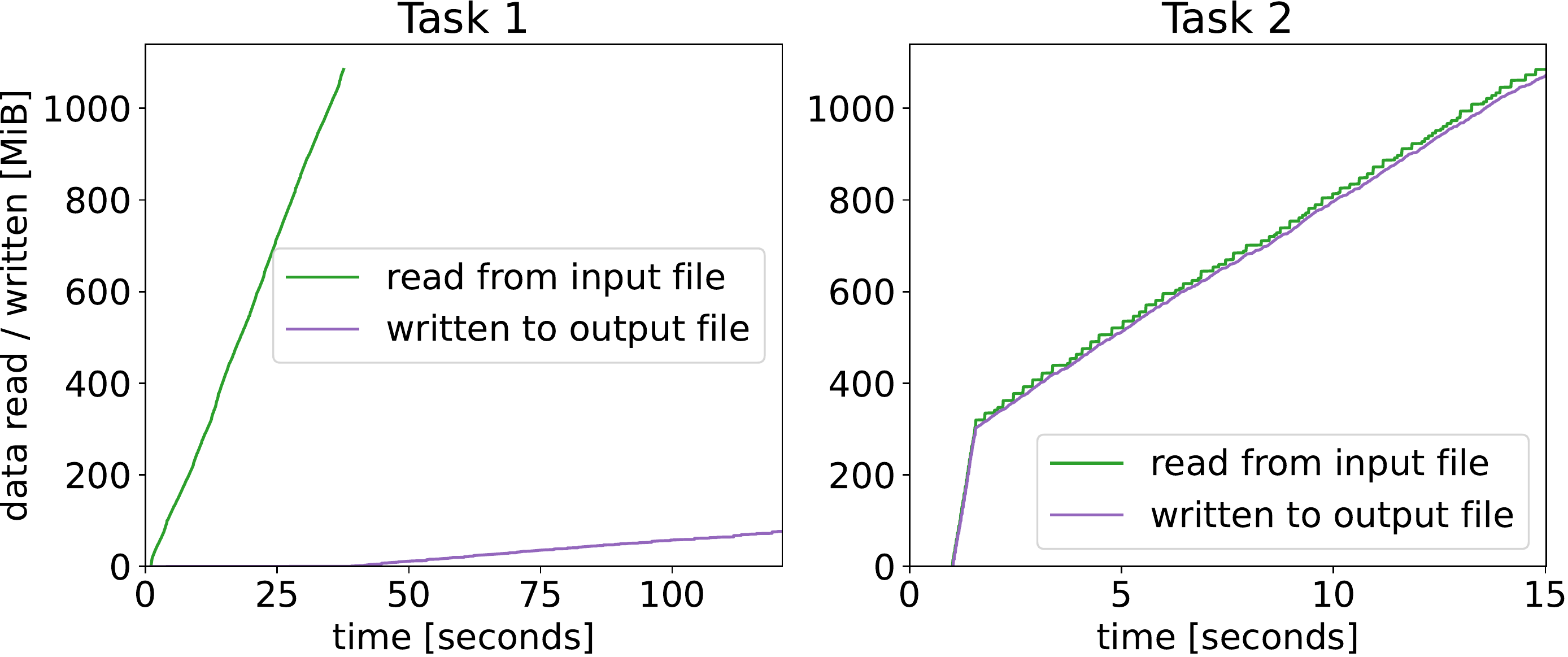}
  \caption{
    Measured I/O activity over time for the isolated execution of evaluation tasks 1 and 2 with input data available on a local disk.
    Task 1 does not start generating output until the whole input has been read.
    Task 2 shows the behavior of a typical streaming task: output and
    input behave similarly. If input rises faster, e.g., because the
    file is still in the cache, as it happened here in the beginning, the output is generated faster as well.
  }
  \label{fig:evaltaskio}
\end{figure}

The workflow was modeled using the proposed framework. Each ffmpeg task is represented by a model process. Additionally, two processes model the download from the webserver preceeding the processes for tasks 1 and 2 respectively, adding up to five processes in total. The download processes depend on one data input, which is the file on the webserver, and one resource, i.e., the data rate.

All requirement functions for data and resources in the evaluation model belong to the common types shown in \figref{fig:samplerequirements}. The more accurate the functions used in the model are, the better the results will reflect the reality and the better the analysis provided by BottleMod will be. As the workflow was developed just for this evaluation, the tasks were investigated and the requirement functions were known well. Similary in real world settings tasks may be annotated by their authors to provide insight about their behavior, maybe even directly stating their requirement functions. Alternatively executions of such tasks can be logged and the requirement functions can be derived from such logs. However, that is part of future work.

The I/O activity of task 1 and task 2 was monitored using BPF~\cite{ebpf} and is visualized in \figref{fig:evaltaskio}. The tasks were executed outside of the workflow with the input data available on a local disk.
The data requirement functions $\mathcal{R}_{Dk}(n_{Dk})$ for all streaming processes are proportional functions with a slope of \emph{outputSize / inputSize}. These are the data requirement functions of the download processes and the processes for tasks 2 and 3, where the latter has two such functions---one per data dependency.
For the process representing task 1, the data requirement function consists of two constant pieces. The first being 0 until (but excluding) the last input byte. The process's \emph{outputSize} is used as a constant function for the last byte, modeling that all progress can be made when all input bytes have been read.

The resource requirement functions $\mathcal{R}_{R\ell}(p)$ are linear functions with only one piece for every process in the evaluation model.
For the processes representing the tasks, we took the CPU time measure in a local execution (in CPU seconds) and spread that evenly among the whole progress. So, the linear function has a slope of \emph{executionTime / outputSize}. For the download processes, the resource requirement function was simply set to a linear function with a slope of $1$, since every output byte needs the link capacity to transfer one byte (e.g., 1\,MByte/s for 1\,ms or 1\,Byte/s for 1 second).

For both download processes, the data input function $I_D(t)$ was constant and equalling the video files size of 1.1\,GB as the file is entirely available on the webserver from the beginning of the workflow execution. The data input functions for the processes representing the tasks are the result functions $O(P(t))$ of their predecessor processes. So, for tasks 1 and 2, the data input functions are the results of their corresponding download processes, and the two data input functions for task 3 are the result functions of tasks 1 and 2, respectively.
As resource input functions $I_R(t)$, the processes representing the tasks got a constant function set to $1$, meaning that in every real-time second, the process could make progress for one CPU second. Along with the resource requirement functions, as specified above, the BottleMod analysis would match the measured execution times if the tasks' processes would be analyzed in isolation and without limiting data inputs.
The process of task 1's download gets assigned the specified portion of the maximum data rate. After analyzing that process, the consumed data rate is set for the process retrospectively. That does not influence the progress as setting the resource input to the actually consumed amount does not introduce any bottleneck. However, it allows assigning the other download process the rest, meaning the difference between the known maximum data rate and the data rate of task 1's download process. Doing that, the cumulatively assigned data rate for the shared link stays at or below its maximum. As maximum data rate, the measured net data rate of 97.51\,Mbit/s is used in the model.

For the evaluation every process only has one output and, therefore, one assigned output function. For every process the identity function $O(p) = p$ was used. The metric for progress is thus the generated amount of output data (in bytes) for these processes.

As task 3 starts when both task 1 and 2 are finished it has its input data completely available from its beginning. Therefore it does not depend on the actual result functions of the processes for task 1 and 2, but only on the time after which both task 1 and 2 are finished. Task 3's analysis starts at that time and the end time denotes the finish of the whole workflow execution.

\subsection{Results}

\begin{figure}[!t]
  \centering
  \includegraphics[width=.635\linewidth]{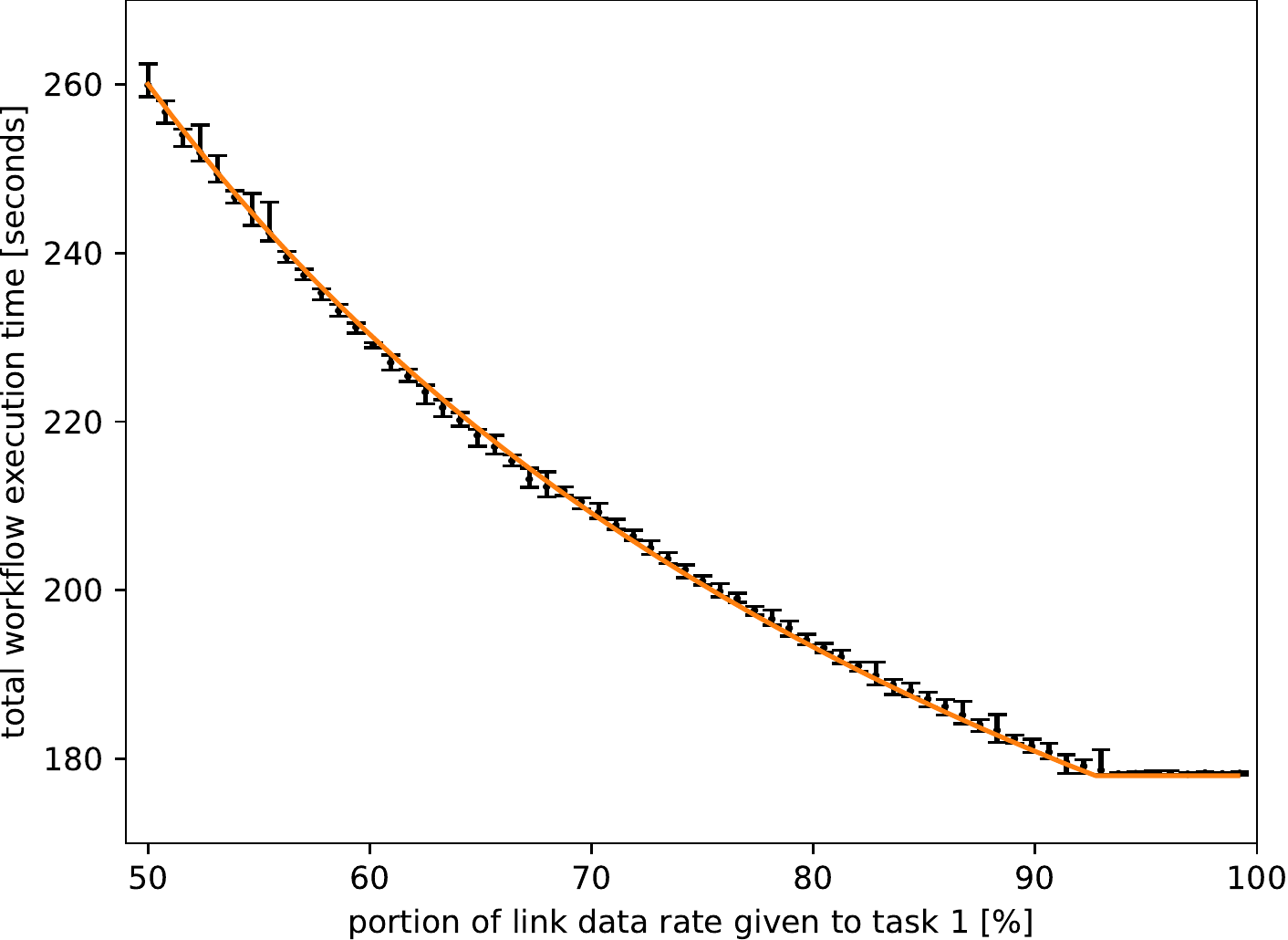}
  \caption{Total execution time in seconds predicted by BottleMod for 600 different prioritizations (orange, solid line). Measured execution time (averages over 10 executions with min/max deviation as bars in black). Different fractions of the link data rate assigned to task 1 (the rest is assigned to task 2) from a total of 100\,Mbit/s.}
  \label{fig:evalcompare}
\end{figure}

\begin{figure}[!t]
  \centering
  \begin{tikzpicture}[textnode/.style={text height=1.5ex,text depth=.25ex}]
    \node[anchor=south west] (fig) at (0,0)
      {\includegraphics[width=.635\linewidth]{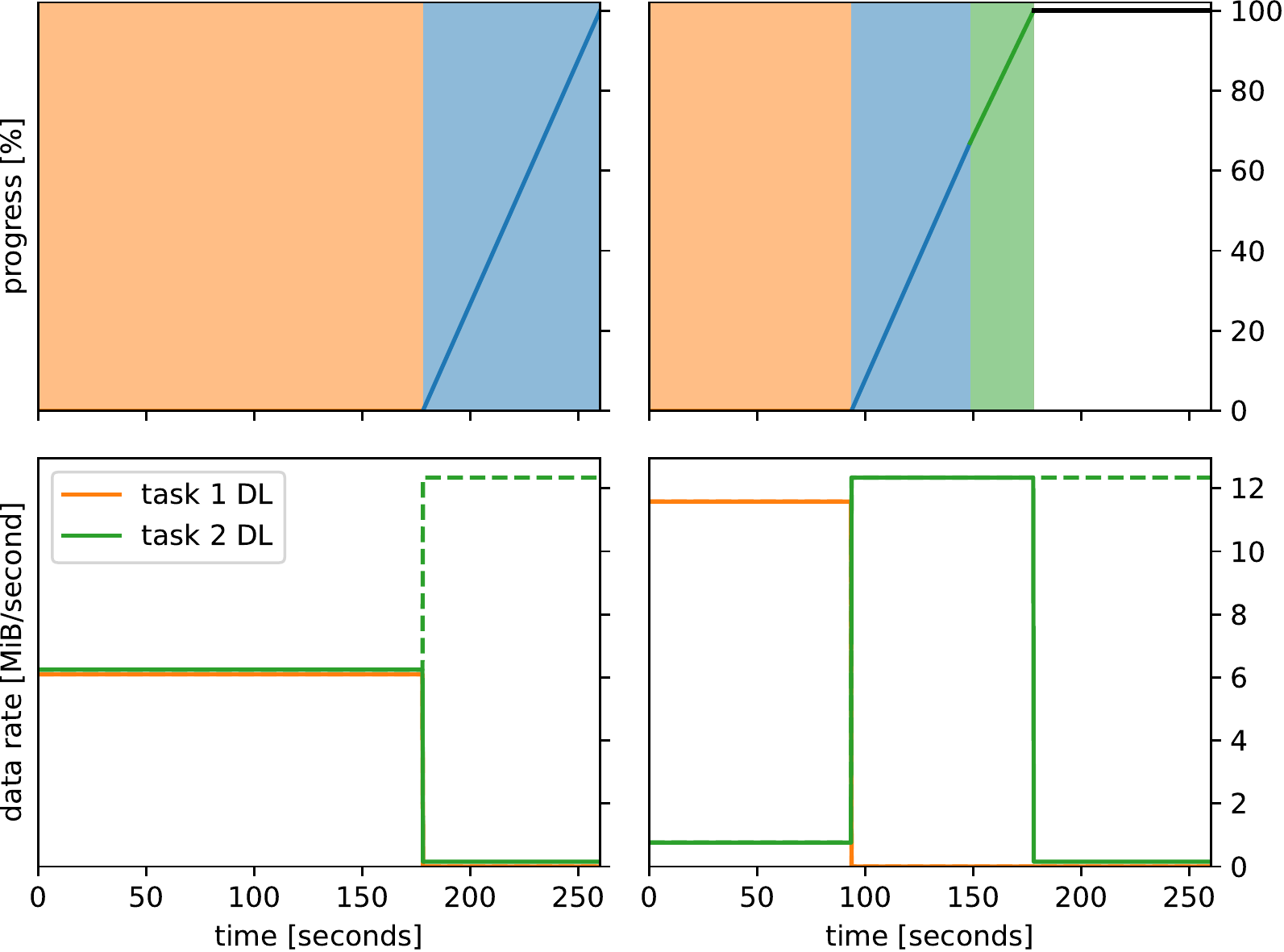}};
      \coordinate (originx) at ($ (fig.south west)!.042!(fig.south east) $);
      \coordinate (originy) at ($ (fig.south west)!.105!(fig.north west) $);
      \coordinate (maxx) at ($ (fig.south west)!.919!(fig.south east) $);
      \coordinate (maxy) at ($ (fig.south west)!.98!(fig.north west) $);

      \coordinate (origin) at (originx |- originy);
      \coordinate (max)    at (maxx |- maxy);
      \path let \p1 = (origin), \p2 = (max) in
      node[anchor=north west] at \graphcoord{-0.005}{1.005}
      {\includegraphics[width=.635\linewidth*\real{.184}]{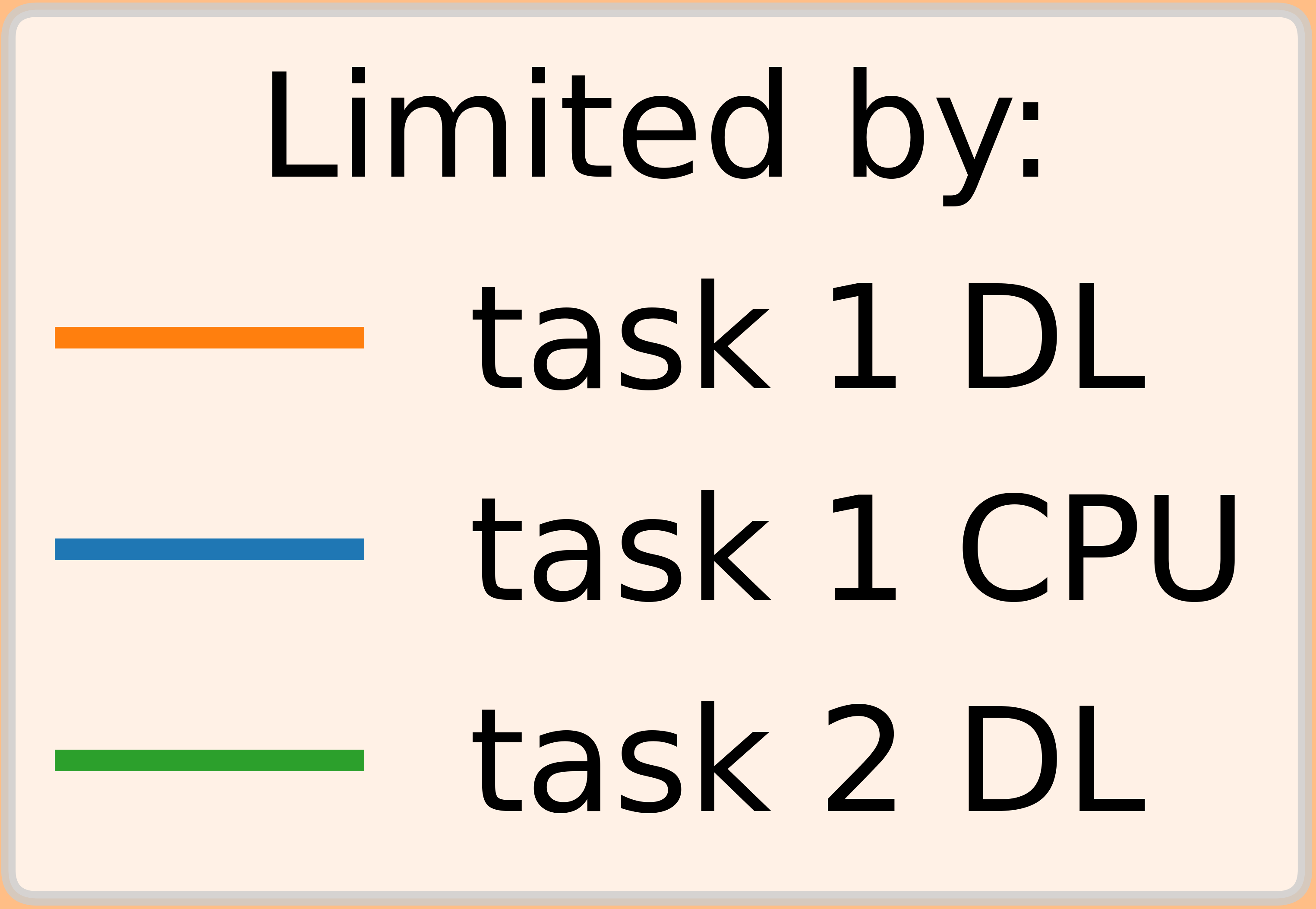}};
  \end{tikzpicture}
  \caption{Combined progress of task 1 and 2 process of \figref{fig:evalworkflow} with bottlenecks (top) and usage of the link data rate (bottom) for the evaluation case with 50\,\% (left) and 95\,\% (right) of the total data rate assigned to task 1's download process. If task 1's download process gets assigned 95\,\% of the data rate from the beginning, task 2's download is slower than the download and processing of task 1, forming an additional bottleneck (green section in the top right graph).}
  \label{fig:evalbottlenecks}
\end{figure}
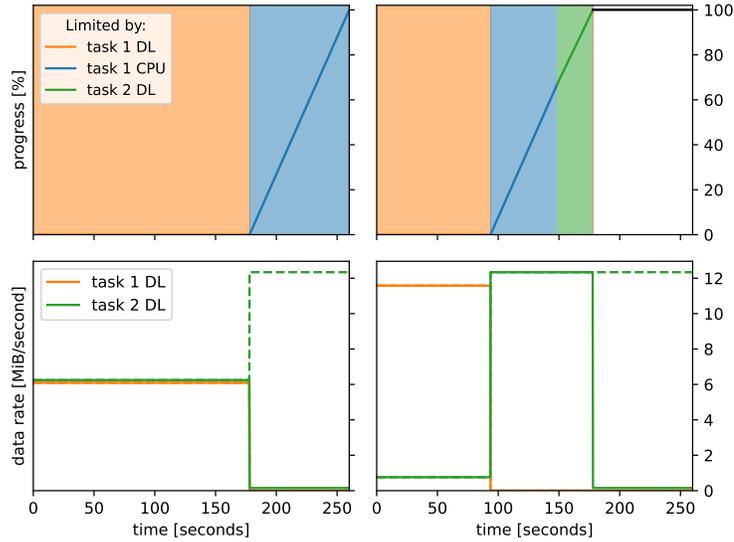

We analyzed the execution with different initial data rate fractions for task 1's download process. For different link fractions, the total execution time predicted by BottleMod is compared to the actual execution time (\figref{fig:evalcompare}). For this model the predicted execution times seem to be well within the range of the actual execution times.
\figref{fig:evalbottlenecks} shows two scenarios in more detail. Here, BottleMod also provides the bottlenecks as well as the data rate assignments during execution. The overall execution time of the workflow, as predicted by BottleMod, is 32\,\% shorter when task 1's download process gets assigned 93\,\% or more of the available data rate instead of just 50\,\%. The latter assignment, dividing the shared resource fairly to the applications that request it, is the default strategy for resource management as there is no specific information about future application behavior. Even if information about the applications would not be as accurate as it is in this scenario, as long as it is somewhat correct, some increase in performance can be expected.

\section{Performance}
\label{sec:performance}

Performance is a crucial property for bottleneck analysis. The more
lightweight but still accurate a framework is, the more often it is
affordable to be executed. Almost instant analysis could
be done periodically during runtime with updated measurements to
steer resource allocation dynamically.  In this section, we
compare the execution time of BottleMod with the state-of-the-art
workflow simulation framework \WRENCH{}~\cite{wrench}.

\WRENCH{} is a framework built on SimGrid~\cite{simgrid} that models and simulates distributed computing applications.  It thus can model and analyze workflow management systems and the coordination of workflows. Tasks and their behavior are not simulated in greater detail~\cite{wrench}. They have inputs and outputs and need a certain amount of computational resources in \WRENCH{}. In contrast to BottleMod, the exact behavior, i.e., when a task needs resources and how much, is not simulated. Therefore, \WRENCH{}'s task model is more basic and less accurate than BottleMod's, especially for long-running tasks that may change behavior over runtime.

A fair performance comparison between BottleMod and \WRENCH{} needs us to simulate the same workflow. Here our exemplary workflow from \figref{fig:evalworkflow} was used again.
Unfortunately, not all workflows can be modeled using \WRENCH{} as it handles tasks as independent execution units and does not directly support data streaming and thus pipelined task execution, which is the default case for BottleMod. In addition, \WRENCH{} can only simulate fairly shared links, and thus data rate limitations on file transfers, as used in our scenario, cannot easily be modeled. Mimicking the data rate limitation in \WRENCH{} by adding a slow disk for certain files does not overcome the limitation, as it is hard to release the limitation when one of the concurrent transfers finished.

For our direct comparison, we dropped the
asymmetric link sharing by just using the 50:50 bandwidth sharing case to overcome \WRENCH{}'s limitations. Thus, both
downloads finish at the same time. Version 2.0 of \WRENCH{} was used, which depends on SimGrid version 3.31. The simulation was executed 1000
times on the host machine that also ran the evaluation (see
above). BottleMod needed 20.0\,ms in that scenario and \WRENCH{} 32.8\,ms per
analysis.
The difference between BottleMod and \WRENCH{} grows a lot when simulating a larger input
file. For BottleMod, that only affects the location where the
mathematical functions change and does not impact the simulation
performance. In contrast, \WRENCH{} simulates more disk reads and network
packet traffic for a larger file, which increases simulation
time. Thus, for an input file size of 100\,GB in the same scenario,
BottleMod only takes 22.8\,ms per simulation while \WRENCH{} needs 1.137 seconds.
There are some differences in the performance of BottleMod as the used methods of SciPy for finding intersections and roots of functions may converge slightly different. However, that is merely noise, the runtime does not scale with the simulated input data size at all. This is fundamentally different to \WRENCH{}.

\section{Related Work}
\label{sec:related-work}

A widely used and well-maintained framework for simulating tasks and workflows is SimGrid~\cite{simgrid}. It is also the base used by \WRENCH{}~\cite{wrench}, which focuses on the simulation of workflows and was compared to BottleMod performancewise in \secref{sec:performance}. The approach is significantly different than the approach described in this paper. SimGrid is a discrete event simulation, comparable to ns3~\cite{ns3} for networks, which also can be integrated into SimGrid. Instead of using mathematical functions to describe a task or process, events are used to express behavior. That may be more detailed and yield more accurate results. But the information about events is more specific and therefore often harder to acquire, especially for tasks that merely are observable from the outside as a black box. On the other hand, given a task's source code, SimGrid can extract its needed events directly from the code with some restrictions. Since these approaches often simulate many fine-grained events, the performance may get worse fast for larger simulated tasks.
SimGrid focuses more on the interaction between tasks and processes. Most examples provided by SimGrid address applications that do very detailed bidirectional communication. Our model focuses more on the tasks execution and their main data flow, rather than on discrete messages exchanged between two tasks. From a data exchange perspective, this approach is broader and less detailed. It closely resembles the acyclic dependency graph of tasks, representing a typical data analysis workflow which is also the context BottleMod was developed in. Each task of a scientific workflow receives input from preceeding tasks and produces an output, which is used in subsequent tasks. Communication and data flow is usually unidirectional for any two concrete workflow tasks and can be easily represented in this model.
Thus, the proposed model does focus more on the result generation pattern of each task. That could also be simulated using SimGrid but would be less efficient.

Predicting resource usage is not a new problem. Quite early processes were clustered by their resource usage from previous executions. These clusters were used to predict resource usage, helping load balancing purposes~\cite{analysis_clustering}. Neural networks were also be used for clustering of processes and prediction~\cite{analysis_nn}.
This approach uses data from many previous executions of many different processes to derive the clusters which are used for the prediction. In contrast BottleMod is able to describe the individual process behavior, not summarizing behavior of many processes into clusters which in turn can provide more accurate predictions. 

Declarative data analysis frameworks such as Spark~\cite{spark} and Flink~\cite{flink} try to assign resources where they are most needed by the concept of lazy evaluation. The required data flow to compute particular outputs is defined, and when the output is needed, the input dependencies are followed backward, and the corresponding computational task is triggered to produce it. Thus, the execution environment only sees the immediately important tasks and assigns resources to them instead of executing the whole data flow from the beginning. That is closer to the reactive model of regular operating system resource managers. Only tasks solely defined in the data flow language might be optimized better. However, that is often not the case for scientific workflows where standalone tools are used that might even be closed source.
Dedicated schedulers like Klink~\cite{klink} optimize output latency of streaming applications with window queries written in such frameworks by tracking the progress of queries using watermarks and prioritizing queries close to their finish. That is closer to a more proactive approach by taking into the account the query's progress, here defined through watermarks. However, while yielding significant improvements with such an approach, it is only applicable for certain stream processing applications and queries and not for general black box applications often used in scientific workflows.
Our approach might help such frameworks, given information about the tasks that need to be executed, by providing the ability to simulate the execution in advance. That way, the framework may assign resources such as CPU time to the processes that benefit most.


Dynamic resource allocation is a challenge in several practical
systems. For example, in the context of web applications sharing the
same hardware with different expectations regarding response time, the
load depends on the user behavior and interest, which makes load
predictions harder and requires direct monitoring to meet the quality-of-service constraints~\cite{weballoc}. In contrast, BottleMod was
developed in the context of scientific workflows. Resource
requirements of a workflow may depend on the input data but are often
well predictable.  Another example is the dynamic resource allocation
for GPGPU tasks under quality-of-service constraints~\cite{gpualloc}.
GPU resources not needed for the quality-of-service constraints of one
application can be used for other applications or be shut down to
save energy. BottleMod does not provide a concrete allocation
algorithm but a static framework to analyze and predict
resource requirements. It allows other components, such as a
scheduler, to base their decisions on those predictions to find efficient
allocations. As BottleMod is fast, it can be repeatedly executed
online with an updated state from monitoring to serve a dynamic resource
allocation algorithm just in time.



\section{Conclusion}

We presented BottleMod, a comprehensive technique to model and analyze the behavior of (chains of) tasks mathematically. It can predict the runtime and resource usage of a process. Additionally, it can be used to find potential execution bottlenecks analytically and provide detailed information about them. This includes identifying the root cause and predicting the performance gain when the bottleneck is remedied.

The precision of BottleMod relies on the granularity of the provided mathematical functions. By differentiation of \emph{input} and \emph{output} functions, there is a seperation of concerns between the execution and the process-specific details of the model. This can enable easier application of the model for task authors as well as execution environment administrators.

The evaluation (\secref{sec:eval}) showed that, given accurate information about the processes and the resource allocation, BottleMod already works very precise. In reality this can be a challenge since extracting accurate data for BottleMod can be difficult. This especially applies for tasks that are considered as black boxes. Their behavior can still be figured out and modeled through extensive logging during exemplary executions using, e.g., extended Berkeley packet filter~\cite{ebpf} (based on the original Berkeley packet filter~\cite{cbpf}). The behavior can also be figured out by running experiments where the task is executed in an environment controlling the needed resources and input data, making them available in a controlled manner.

The results of BottleMod contain the progress, bottlenecks and predicted resource usage of a process. All this information can be used by a resource manager to schedule resources proactively. The results could also enable more efficient placement of tasks, running as part of workflows in distributed environments, on nodes. In order to optimize resource allocation and mitigate the most disadvantageous bottlenecks, it is up to the resource manager to apply the insights found by the BottleMod analysis to the actual execution. The resource manager could change priorities or limit resources with methods such as Linux control groups~\cite{cgroups} or SDN bandwidth reservations.

The practical implementation is able to represent functions by piecewise polynomials. Calculation of the results only needs to iterate over few locations of the progress function, namely the points where the current bottleneck or a relevant functions piece changes. This behavior is fundamentally different to existing approaches based on simulation of discrete events. Depending on the number of events using BottleMod can be much faster. This advantage can even be further increased by using, e.g., piecewise linear functions instead of polynomials. Due to this efficient analysis, BottleMod may even be used while the tasks or the workflow is still executing to conduct certain optimizations just in time, incorporating the most recent information to deliver even more precise results.

Our BottleMod implementation the scripts used to execute our evaluation will be made publically available until (no later than) submission of the camera-ready paper under: \url{https://github.com/bottlemod/bottlemod}.

\bibliographystyle{ACM-Reference-Format}
\bibliography{literature}

\appendix
\section{Evaluation Workflow}
\label{sec:appendix}

\noindent
Task 1 commands:

\noindent{\scriptsize
\texttt{\# create pipe}\\[-1mm]
\texttt{mkfifo pipe1}\\[-1mm]
\texttt{\# set the buffersize of the pipe to 256 MiB (own small C program using fcntl() to set F\_SETPIPE\_SZ)}\\[-1mm]
\texttt{./setpipesize pipe1 \$((256 * 1024 * 1024)) \&}\\[-1mm]
\texttt{\# set maximum channel bandwidth for task 1 (using port 8080)}\\[-1mm]
\texttt{nft add rule inet filter input ip protocol tcp ip saddr 192.168.75.142 tcp sport   8080~$\backslash$\\[-1mm]
\mbox{}~~limit rate over \$RATE\_TASK1 kbytes/second drop
\\[-1mm]
\texttt{\# reverse the video (in background), reading from the pipe blocks until data arrives}\\[-1mm]
ffmpeg -y -to 05:00 -f mp4 -i pipe1 -an -vf reverse -c:v libx264 -preset veryfast -threads 2 -movflags~$\backslash$\\[-1mm]
\mbox{}~~frag\_keyframe+empty\_moov tmp/t1.mp4 \&
\\[-1mm]
\texttt{\# download the video file, writing it to the pipe}\\[-1mm]
wget -O pipe1 http://192.168.75.142:8080/video.mp4
\\[-1mm]
\texttt{\# after download finished, release the full bandwidth to the other task}\\[-1mm]
nft replace rule inet filter input handle 2 ip protocol tcp ip saddr 192.168.75.142 tcp sport 80~$\backslash$\\[-1mm]
\mbox{}~~limit rate over \$RATE\_TOTAL kbytes/second drop}}

\medskip
\noindent
Task 2 commands:

\noindent{\scriptsize
\texttt{\# set maximum channel bandwidth for task 2 (using port 80)}\\[-1mm]
  \texttt{nft add rule inet filter input ip protocol tcp ip saddr 192.168.75.142 tcp sport   80~$\backslash$\\[-1mm]
    \mbox{}~~limit rate over \$RATE\_TASK2 kbytes/second drop
    \\[-1mm]
\texttt{\# rotate the video}\\[-1mm]
    ffmpeg -y -i http://192.168.75.142:80/video.mp4 -an -c:v copy -metadata:s:v:0 rotate=180 -movflags~$\backslash$\\[-1mm]
    \mbox{}~~frag\_keyframe+empty\_moov tmp/t2.mp4
    \\[-1mm]
\texttt{\# when finished, release the full bandwidth to the other task}\\[-1mm]
    nft replace rule inet filter input handle 3 ip protocol tcp ip saddr 192.168.75.142 tcp sport 8080~$\backslash$\\[-1mm]
    \mbox{}~~limit rate over \$RATE\_TOTAL kbytes/second drop}}

\medskip
\noindent
Task 3 command:

\noindent{\scriptsize
\texttt{\# combine video outputs of task 1 and task 2}\\[-1mm]
\texttt{ffmpeg -y -to 05:00 -i tmp/t1.mp4 -to 05:00 -i tmp/t2.mp4 -map 0 -map 1 -c copy result.mp4}}

\end{document}